\documentclass[a4paper,UKenglish]{lipics-v2018}
\usepackage{microtype}%if unwanted, comment out or use option "draft"
\usepackage{times}
\usepackage{xcolor}
\usepackage{soul}
\usepackage[utf8]{inputenc}
\usepackage{amsmath}
\usepackage{amssymb}
\usepackage{amsthm}
\usepackage{stmaryrd}
\usepackage{graphicx}
\usepackage{subcaption}
\usepackage{booktabs}
\usepackage{verbatim}
\usepackage[ruled, linesnumbered, vlined]{algorithm2e} % For algorithms

\newtheorem*{theorem*}{Theorem}
\newtheorem*{proposition*}{Proposition}
\newtheorem*{lemma*}{Lemma}

\def\dfar{$\mathit{DFA_{R}}$}
\def\nfar{$\mathit{NFA_{R}}$}
\def\dfasr{$\mathit{DFA_{\Sigma^{*}\cdot R}}$}

\def\pmcmr{$\mathit{PMC_{R}^{m}}$}
\def\pmczeror{$\mathit{PMC_{R}^{0}}$}
\def\pmconer{$\mathit{PMC_{R}^{1}}$}
\def\pfc{$\mathit{P_{fc}}$}

\DeclareMathOperator*{\argmin}{arg\,min}

\title{Event Forecasting with Pattern Markov Chains}

\titlerunning{Event Forecasting with Pattern Markov Chains}%optional, please use if title is longer than one line

\author{Elias Alevizos}{National Center for Scientific Research (NCSR) ``Demokritos'', Greece}{alevizos.elias@iit.demokritos.gr}{}{}%mandatory, please use full name; only 1 author per \author macro; first two parameters are mandatory, other parameters can be empty.

\author{Alexander Artikis}{University of Piraeus, Greece\\National Center for Scientific Research (NCSR) ``Demokritos'', Greece}{a.artikis@unipi.gr}{}{}

\author{Georgios Paliouras}{National Center for Scientific Research (NCSR) ``Demokritos'', Greece}{paliourg@iit.demokritos.gr}{}{}

\authorrunning{E. Alevizos et. al.}%mandatory. First: Use abbreviated first/middle names. Second (only in severe cases): Use first author plus 'et. al.'

\Copyright{Elias Alevizos, Alexander Artikis and Georgios Paliouras}%mandatory, please use full first names. LIPIcs license is "CC-BY";  http://creativecommons.org/licenses/by/3.0/

\subjclass{
\ccsdesc[300]{Theory of computation~Formal languages and automata theory},
\ccsdesc[300]{Theory of computation~Pattern matching},
\ccsdesc[300]{Theory of computation~Random walks and Markov chains},
\ccsdesc[500]{Information systems~Data streaming}
}% mandatory: Please choose ACM 2012 classifications from https://www.acm.org/publications/class-2012 or https://dl.acm.org/ccs/ccs_flat.cfm . E.g., cite as "General and reference $\rightarrow$ General literature" or \ccsdesc[100]{General and reference~General literature}. 

\keywords{Complex event processing, Stream processing, Finite Automata, Regular Expressions}%mandatory

\category{}%optional, e.g. invited paper

\relatedversion{}%optional, e.g. full version hosted on arXiv, HAL, or other respository/website

\supplement{}%optional, e.g. related research data, source code, ... hosted on a repository like zenodo, figshare, GitHub, ...

\nolinenumbers %uncomment to disable line numbering
\hideLIPIcs  %uncomment to remove references to LIPIcs series (logo, DOI, ...), e.g. when preparing a pre-final version to be uploaded to arXiv or another public repository
\begin{document}

\maketitle

\begin{abstract}
We present a system for online probabilistic event forecasting. 
We assume that a user is interested in detecting and forecasting event patterns,
given in the form of regular expressions.
Our system can consume streams of events and forecast when the pattern is expected to be fully matched.
As more events are consumed, the system revises its forecasts to reflect possible changes
in the state of the pattern.
The framework of Pattern Markov Chains is used in order to learn a probabilistic model for the
pattern, with which forecasts with guaranteed precision may be produced, 
in the form of intervals within which a full match is expected.
Experimental results from real-world datasets are shown and the quality of the produced forecasts is explored, using both precision scores and two other metrics:
spread, which refers to the ``focusing resolution'' of a forecast (interval length), and distance, which captures how early a forecast is reported.
\end{abstract}

\section{Introduction}

As analytics moves towards a model of proactive computing,
the requirement for forecasting acquires more importance \cite{etzion_proactive_2016}.
Systems with forecasting capabilities can play a significant role in assisting users to make smart decisions as soon as critical situations are detected.
As an example, consider credit card fraud management.
Automated fraud detection works with patterns consisting of long sequences of transactions with specific characteristics. 
Being able to forecast that part(s) of such sequences have high probability of leading to a full match (i.e., a fraud) can help an analyst focus on the involved cards and possibly take a proactive action even before the system detects the fraud.

The need for event forecasting as a means for proactive behavior has led to proposals about how forecasting could be conceptualized and integrated within a complex event processing system. However, such proposals still remain largely at a conceptual level, without providing concrete algorithms \cite{fulop_predictive_2012,engel_towards_2011}.
On the other hand, there is a substantial body of work on the related field of time-series forecasting \cite{montgomery_introduction_2015}.
However, time-series analysis is usually applied on numerical data streams, where each element of the stream corresponds to a measurement of some variable of interest. 
Moreover, these measurements are often assumed to take place at time intervals of constant length.
On the contrary, event processing systems need to be able to additionally deal with symbolic/categorical streams, where each element might be accompanied by arguments, either numerical or symbolic, arriving at unspecified timepoints.

We present an implementation of a system for event forecasting.
We assume that event patterns are defined through regular expressions. 
As a first step, these patterns are converted to finite automata for the purpose of pattern matching. 
Subsequently, these automata are converted into Markov chains, which allow for the construction of a probabilistic model for the initial pattern. 
The final goal is to be able to forecast, as events arrive at the system,
when the pattern will be fully matched.
This is the first time that Pattern Markov Chains are used for online event forecasting. 
We show that our system can indeed forecast the completion of patterns in real-world datasets and that,
under certain assumptions, 
it can do so with guaranteed precision.
Moreover, we explore the quality of the produced forecasts, using three different metrics:
precision score, 
spread, which refers to how focused a forecast is, 
and distance, which captures how early a forecast is reported.

The structure of the paper is as follows:
Section \ref{sec:related} presents related work.
In Section \ref{sec:theory}, the necessary mathematical terminology and framework are described.
Section \ref{sec:implementation} elaborates on the implementation details of the system,
while Section \ref{sec:experiments} presents experimental results on real-world datasets.
Finally, in Section \ref{sec:summary} we conclude with a summary and a discussion on future directions of research.

\section{Related work}
\label{sec:related}

\begin{table*}[!ht]
\scriptsize
%\small
\setlength{\tabcolsep}{2pt}
\centering
\caption{Methods for event forecasting (in order of publication date).}
\begin{tabular}{l*{5}{c}} 
\toprule
\textbf{Paper}  &  \textbf{Prob.} & \textbf{Learning} & \textbf{Language} & \textbf{Relations} & \textbf{Focus} \\ 
\midrule
\cite{ghallab_chronicles_1996} & no & \parbox{3.0cm}{\centering Syntactical pattern recognition} & \parbox{3.0cm}{\centering Chronicles ($\mathit{before}$, $\mathit{equal}$, $\mathit{after}$ + numerical time constraints)} & \parbox{3.5cm}{\centering Only on timestamps} & \parbox{2.0cm}{\centering Expected events following a partial match} \\
\midrule
\cite{weiss_learning_1998} & no & Genetic & \parbox{3.0cm}{\centering Sequences (+ OR operator and wildcard events)} & \parbox{3.5cm}{\centering Discrete arguments. Equality and ``don't care'' operators on single events (no relations between different events). } & \parbox{2.0cm}{\centering Rare events} \\
\midrule
\cite{domeniconi_classification_2002} & no & \parbox{3.0cm}{\centering Singular Value Decomposition + Support Vector Machines} & Feature matrix & \parbox{3.5cm}{\centering Discrete arguments. Equality on single events (no relations between different events).} & \parbox{2.0cm}{\centering Rare events} \\
\midrule
\cite{vilalta_predicting_2002} & no & \parbox{3.0cm}{\centering Association (predictive) rule mining} & \parbox{3.0cm}{\centering Sets of events (not necessarily in sequential order)} & no & \parbox{2.0cm}{\centering Rare events} \\
\midrule
\cite{laxman_stream_2008} & yes & \parbox{3.0cm}{\centering Frequent episode discovery + Expectation Maximization} & Sequences & no & Immediately next event \\
\midrule
\cite{xing_mining_2008} & no & \parbox{3.0cm}{\centering Decision trees} & Sequences & no & Class labels of sequences \\
\midrule
\cite{cho_online_2010} & no & no & \parbox{3.0cm}{\centering Directed Acyclic Graphs} & \parbox{3.5cm}{\centering (In)equality on single events (no relations between different events)} & Minimal occurrences \\
\midrule
\cite{muthusamy_predictive_2010} & yes & \parbox{3.0cm}{\centering Learns probabilistic model. No pattern mining.} & \parbox{3.0cm}{\centering Sequences (+ $\mathit{conjunction}$ and $\mathit{disjunction}$)} & \parbox{3.5cm}{\centering yes} & \parbox{3.0cm}{\centering Completion time of pattern} \\
\midrule
\cite{gunawardana_model_2011} & yes & \parbox{3.0cm}{\centering Decision trees + Conditional Intensity Models} & Sequences & no & Event sequences \\
\midrule
\cite{fahed_efficient_2014} & no & \parbox{3.0cm}{\centering Frequent episode discovery (starts from the consequent)} & Sequences & no & \parbox{3.0cm}{\centering Minimal antecedent, distant consequent}\\
\midrule
\cite{zhou_pattern_2015} & no & \parbox{3.0cm}{\centering Sequential pattern mining} & Sequences  & no & \parbox{3.0cm}{\centering Online update of patterns} \\
\midrule
Our approach & yes & \parbox{3.0cm}{\centering Learns probabilistic model. No pattern mining.} & Regular expressions  & no & \parbox{3.0cm}{\centering Completion time of pattern} \\
\bottomrule
\end{tabular}
\label{tab:related}
\end{table*}

Timewewaver is a genetic algorithm that tries to learn from sequences of events a set of predictive patterns \cite{weiss_learning_1998}. Its focus is on learning patterns that can forecast, 
as early as possible, rare events, such as equipment failures.
In \cite{domeniconi_classification_2002}, the forecasting problem is formulated as a classification problem and the goal is again to construct predictive patterns for 
rare events. 
The proposed algorithm constructs a matrix of features, finds a reduced set of features through Singular Value Decomposition and then trains a set of Support Vector Machines, one for each target event.

A significant number of forecasting methods comes from the field of temporal pattern mining, where patterns are usually defined either as association rules \cite{agrawal_mining_1993} or as frequent episodes \cite{mannila_discovery_1997}.
For example, in \cite{vilalta_predicting_2002}, a framework similar to that of association rule mining is used in order to identify sets of event types that frequently precede a rare, target event within a temporal window. 
In \cite{laxman_stream_2008}, a probabilistic model is presented.
The goal is to calculate the probability of the immediately next event in the stream through a combination of standard frequent episode discovery algorithms, Hidden Markov Models and mixture models.
Episode rules constitute the framework of \cite{fahed_efficient_2014} as well,
where the goal is to mine predictive rules whose antecedent is minimal (in number of events) and temporally distant from the consequent.
The algorithms presented in \cite{zhou_pattern_2015} focus on batch, online mining of sequential patterns, without maintaining exact frequency counts.
At any time, the learned patterns (up to that time) can be used to test whether a prefix matches the last events seen in the stream and therefore make a forecast.

In \cite{xing_mining_2008}, a variant of decision trees is used in order to learn
sequence prefixes that are as short as possible and that can forecast the class label
of the whole sequence. 
The method proposed in \cite{cho_online_2010} starts with a given episode rule (as a Directed Acyclic Graph) and
the goal is to build appropriate data structures that can efficiently detect the
minimal occurrences of the antecedent of a rule defining a complex event, 
i.e., those ``clusters'' of antecedent events that are closer together in time.
In \cite{gunawardana_model_2011}, Piecewise-Constant Conditional Intensity Models and decision trees are employed in order to learn a very fine-grained model of the temporal dependencies among events in sequences. 
The learned models can then be used to calculate whether a sequence of target events will occur in a given order and in given time intervals.
One of the earliest methods for forecasting is the Chronicle Recognition System, proposed in \cite{ghallab_chronicles_1996,dousson_chronicle_2007},
where events may be associated with both temporal operators and with numerical constraints on their timestamps. 
The system uses partial matches in order to
report when the remaining events are expected for the pattern to complete.
However, such forecasts are not based on a confidence or probability metric. 

The work most closely related to ours is the one presented in \cite{muthusamy_predictive_2010},
where Markov chains are also used in order to estimate when a pattern is expected to be fully matched.
This work can also handle some relational constraints. 
On the other hand, the framework of Pattern Markov Chains that we use offers three main advantages:
first, it can handle arbitrary regular expressions (and not only sequential patterns);
second, it can handle streams generated by higher-order processes;
third, it automatically calculates the expected time interval of pattern completion.

Table \ref{tab:related} summarizes the methods presented in this section.
Its second column (\textit{Prob.}) indicates whether a method employs a probabilistic framework. 
The third column (\textit{Learning}) shows whether (and how) a method can automatically extract such patterns by reading (part of) the input event stream.
The next two columns (\textit{Language} and \textit{Relations}) refer to the expressivity of the (learned or given) patterns.
The entries in the \textit{Language} column show how the different events in a pattern may be temporally related. 
The \textit{Relations} column mentions if (and how) arguments of the involved events may be constrained and related.
Finally, the goal of the last column is to show on what kind of forecasts each method focuses. 

The last row shows how our approach compares to other methods.
The advantage of our approach is that it moves beyond simple sequential patterns and
combines the expressive power of regular expressions with a rigorous probabilistic framework.
The focus in this paper is on estimating when a full match of a given pattern will be detected.
Note, however, that this does not exclude the possibility of incorporating (some of) the extra functionality of other methods.
For example, frequent pattern mining could be a possible future extension by using the theory of Markov chains in order to estimate the expected number of occurrences of a pattern.
One of the most important challenges for all methods (including ours) is \textit{relationality},
i.e., the ability to handle patterns in which the arguments of \textit{different} events in the pattern are related through some constraints.

\section{Theoretical background}
\label{sec:theory}

The problem we address could be stated as follows.
Given a stream of events $S$ and a pattern $R$, the goal is two-fold.
First: find the full matches of $R$ in $S$. 
Second: as the stream is consumed by the engine, forecast the full matches before they are detected by the recognition engine. 
For the recognition task, we use finite automata, whereas for forecasting,
we convert these automata into Markov chains.

We make the following assumptions:
\begin{itemize}
\itemsep0em
\item Patterns are defined in the form of regular expressions.
\item The \textit{selection strategy} is either that of \textit{contiguity} (i.e., events in a match must be contiguous, without irrelevant events intervening) or \textit{partition-contiguity} (i.e., same as \textit{contiguity} but stream may be partitioned by a specific event attribute). The \textit{counting policy} is that of \textit{non-overlap} (i.e., after a full match, the automaton returns to its start state). 
%See Appendix \ref{sec:count_pol} for a brief discussion. For more extensive discussions, see
See \cite{zhang_complexity_2014} for the various selection strategies and \cite{lladser_multiple_2007} for the counting policies. 
\item The stream is generated by a $m$-order Markov process.
\item The stream is stationary, i.e., its statistical properties remain the same. Hence the constructed Markov chain is \textit{homogeneous} and its transition matrix remains the same at all time-points.
\item A forecast reports for how many ``points'' we will have to wait until a full match. By the term ``point'', we refer to number of transitions of the Markov chain
(or equivalently to number of future events) and not to time-points.
Points are indeed time-points only in cases where a new event arrives at each time-point.
\end{itemize}

The theoretical tools presented in this section are based mostly on the work described in \cite{nuel_pattern_2008,nicodeme_motif_2002,fu_distribution_2003} and are grounded in the field of string pattern matching. 
For a review of this field, the reader may consult \cite{lladser_multiple_2007}. 
Comprehensive treatments of this subject may be found in 
\cite{crochemore_algorithms_2007,gusfield_algorithms_1997,lothaire_applied_2005}.
%\cite{gusfield_algorithms_1997,lothaire_applied_2005}.

\subsection{Event Recognition}
In this section, we briefly review some of the necessary terminology \cite{hopcroft_introduction_2007}.
Regular expressions define the so-called regular languages. Within the context of the theory of regular languages, an \textit{alphabet} $\Sigma=\{e_{1},...,e_{r}\}$ is a finite, non-empty set of symbols. 
The alphabet essentially refers to the set of the different event types that may appear in the stream. 
A string over $\Sigma$ is a finite sequence of symbols from the alphabet. A language $L$ over $\Sigma$ is a set of strings over $\Sigma$. 
One common way to denote languages over $\Sigma$ is through the use of regular expressions. If $R$ denotes a regular expression, then $L(R)$ denotes the language defined by $R$. 
There are three operators that are used in regular expressions: \textit{union}, which is binary and is denoted by the symbol $+$, \textit{concatenation}, again binary, denoted by $\cdot$, and \textit{star closure}, which is unary, denoted by $^{*}$.
Regular expressions are inductively defined as follows:
\begin{itemize}
\itemsep0em
\item The \textit{union} of two languages $L$ and $M$, $L\cup M$ is the set of strings that belong either to $L$ or $M$. If $R_1$ and $R_2$ are regular expressions, then $R_1 + R_2$ is also a regular expression and $L(R_1 + R_2)=L(R_1) \cup L(R_2)$. Union corresponds to the $\mathit{OR}$ operator in event recognition.
\item The \textit{concatenation} of two languages $L$ and $M$, $L \cdot M$ is the set of strings formed by concatenating strings from $L$ with strings from $M$, i.e., 
$L \cdot M = \{s_1\cdot s_2, s_1 \in L, s_2 \in M\}$. If $R_1$ and $R_2$ are regular expressions, then $R_1 \cdot R_2$ is also a regular expression and $L(R_1 \cdot R_2) = L(R_1) \cdot L(R_2)$. 
Concatenation corresponds to the $\mathit{sequence}$ operator in event recognition.
\item The \textit{star closure} of a language $L$ is 
$L^{*} = \bigcup\limits_{i \geq 0}{L^{i}}$, where $L^{i}$ is concatenation of $L$ with itself $i$ times. 
If $R$ is a regular expression, then
$R^{*}$ is also a regular expression and $L(R^{*})=(L(R))^{*}$.
Star closure corresponds to the $\mathit{iteration}$ operator in event recognition.
\end{itemize}
Finally, the inductive basis for a regular expression is that it may also be the empty string or a symbol from $\Sigma$.

Regular expressions may be encoded by deterministic and non-deterministic finite automata (DFA and NFA respectively). 
%and are equivalent to non-deterministic and deterministic finite automata (NFA and DFA respectively) in the sense that they define exactly those languages accepted by NFA and DFA.
%\begin{definition}[Deterministic Finite Automaton]
%\textit{
A DFA is 5-tuple $A=(Q,\Sigma,\delta,q_{0},F)$ where 
$Q$ is a finite set of states, $\Sigma$ a finite set of symbols, 
$\delta: Q \times \Sigma \rightarrow Q$ a transition function from a state reading a single symbol to another state, $q_{0} \in Q$ a start state and $F \subset Q$ a set of final states. A string $s=e_{1}e_{2}...e_{d} \in \Sigma^{*}$ is accepted by the DFA if $\delta(q_{0},s) \in F$, where the transition function for a string is defined as $\delta(q,e_{1}e_{2}...e_{d})=\delta(\delta(q,e_{1}e_{2}...e_{d-1}),e_{d})$.
%}
%\end{definition}
The definition for a NFA is similar with the modification that the transition function is now 
$\delta: Q \times \Sigma \rightarrow S_{Q}$, where $S_{Q}$ is the power set of $Q$.
%, i.e., from a single state, after reading one symbol, there might be more than one next states (hence the non-determinism).

\begin{figure}[t]
    \centering
    \begin{subfigure}[b]{0.45\textwidth}
        \includegraphics[width=\textwidth]{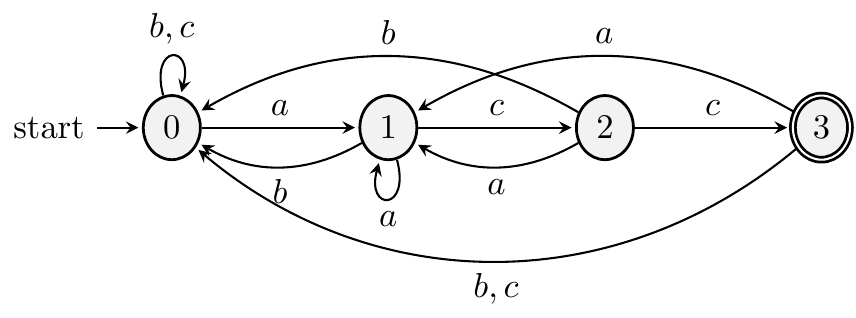}
        \caption{\dfasr.}\label{fig:dfatcc}
    \end{subfigure}
    \hfill
    \begin{subfigure}[b]{0.45\textwidth}
        \includegraphics[width=\textwidth]{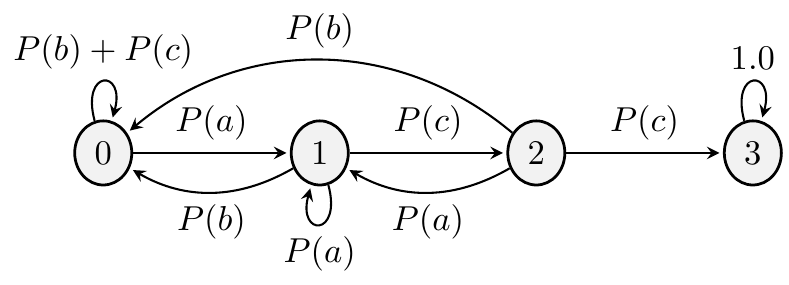}
        \caption{\pmczeror.}\label{fig:mctcc0}
    \end{subfigure}
    \hfill
    \begin{subfigure}[b]{0.55\textwidth}
        \includegraphics[width=\textwidth]{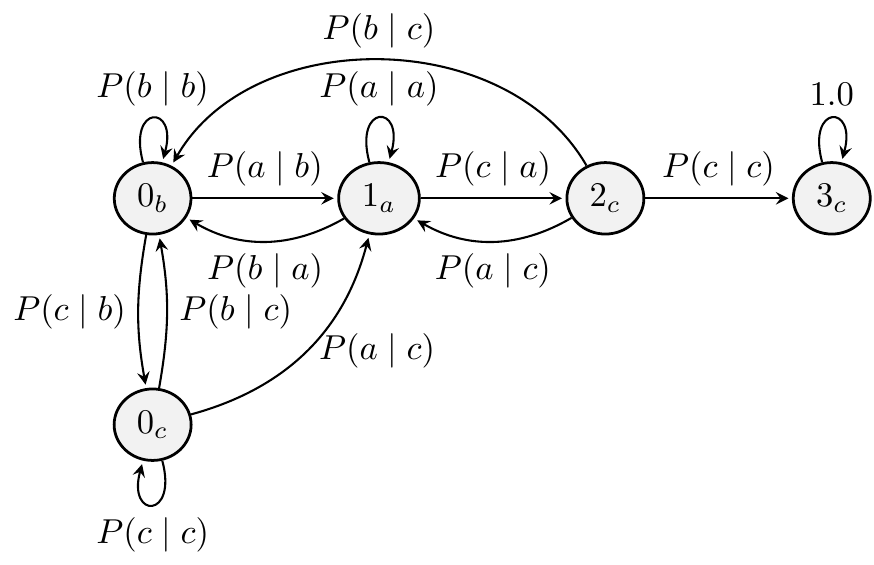}
        \caption{\pmconer.}\label{fig:mctcc1}
    \end{subfigure}
    \caption{DFA and PMCs for $R=a\cdot c\cdot c$,  $\Sigma=\{a,b,c\}$ and for $m=0$ and $m=1$.}\label{fig:dfa_mc_example}
\end{figure}

There exist well-known algorithms for converting a regular expression $R$ to an equivalent NFA, \nfar, and subsequently to an equivalent DFA, \dfar$\ $ \cite{hopcroft_introduction_2007}.
For event recognition,
a slight modification is required so that the DFA can detect all the full matches in the stream.
The regular expression and DFA that should be used are $\mathit{\Sigma^{*} \cdot R}$ and \dfasr$\ $ respectively so that the DFA may recognize all the strings ending with $R$ \cite{nicodeme_motif_2002,crochemore_algorithms_2007,gusfield_algorithms_1997}.
%\cite{nicodeme_motif_2002,gusfield_algorithms_1997}.
Figure \ref{fig:dfatcc} shows an example of a DFA, 
constructed for $R=a\cdot c\cdot c$ (one event of type $a$ followed by two events of type $c$) 
and $\Sigma=\{a,b,c\}$ (three event types may be encountered, $a$,$b$ and $c$).

\subsection{Event Forecasting}

We use Pattern Markov Chains, 
i.e., convert \dfasr$\ $ to an ``appropriate'' Markov chain.
The Markov chain should be ``appropriate'' in the sense that it could be used in order to make probabilistic inferences about the run-time behavior of \dfasr .
In the case where the stream consumed by \dfasr$\ $ is assumed to be composed of
a sequence of independent, identically distributed (i.i.d.) events from $\Sigma$,
then constructing the corresponding Markov chain is straightforward. 
As shown in \cite{nuel_pattern_2008},
if $X=X_{1},X_{2},...,X_{i},...$ is the i.i.d.\ sequence of input events,
then the sequence $Y=Y_{0},Y_{1},...,Y_{i},...$, where $Y_{0}=q_{0}$ and $Y_{i}=\delta(Y_{i-1},X_{i})$
(i.e., the sequence of the states that \dfasr$\ $ visits)  
is a 1-order Markov chain.
Such a Markov chain, associated with a pattern $R$, is called a Pattern Markov Chain (PMC).
Moreover, the transition probabilities between two states are simply given by the occurrence probabilities of the event types. 
If $p,q \in Q$ and $\boldsymbol{\Pi}$ is the $\lvert Q \rvert \times \lvert Q \rvert$ matrix holding these probabilities, 
then $\boldsymbol{\Pi}(p,q){=}P(X_{i}{=}e)$, if $\delta(p,e)=q$ (otherwise, it is $0$).
This means that we can directly map the states of \dfasr$\ $ to the states of a
PMC and for each edge of \dfasr$\ $ labeled with
$e \in \Sigma$, we can insert a transition in the PMC with probability $P(e)$
(assuming here stationarity, i.e., $P(X_{i}=e)=P(X_{j}=e), \forall i,j$).
As an example, Figure \ref{fig:mctcc0} shows the PMC constructed for $R=a\cdot c\cdot c$,
based on the \dfasr$\ $ of Figure \ref{fig:dfatcc}
(the reason why state 3 has only a self-loop with probability $1.0$ will be explained later).

For the more general case where the process generating the stream is of a higher order
$m \geq 1$,
the states of the PMC should be able to remember the past
$m$ symbols so that the correct conditional probabilities may be assigned to its transitions.
However, the states of \dfasr$\ $ do not hold this information.
As shown in \cite{nicodeme_motif_2002,nuel_pattern_2008}
we can overcome this problem by iteratively duplicating those states of \dfasr$\ $
for which we cannot unambiguously determine the last $m$ symbols that can lead to them
and then convert it to a PMC.
From now on,
we will use the notation \pmcmr$\ $ to refer to the Pattern Markov Chain of a pattern $R$ and order $m$.
Please, note that, from a mathematical point of view,
the resulting Markov chain is always of order 1,
regardless of the value of $m$ \cite{nuel_pattern_2008}.

As an example, see Figure \ref{fig:mctcc1} which shows the resulting \pmconer$\ $ for $R=a\cdot c\cdot c$. 
Note that the DFA for the same pattern with $m{=}0$,
shown in Figure \ref{fig:dfatcc},
has a state which is ambiguous.
When in state $0$, the last symbol read may be either $b$ or $c$.
For all the other states, we know the symbol that led to them.
Therefore, state $0$ must be duplicated
and state $0_{c}$ is added.
Now, when in state $0_{b}$, we know that the last symbol was $b$,
whereas in state $0_{c}$, it was $c$.

Once we have \pmcmr$\ $ for a user-defined pattern $R$,
we may use the whole arsenal of Markov chain theory to make certain probabilistic inferences
about $R$. 
%In the works that we have cited in this section \cite{nuel_pattern_2008,nicodeme_motif_2002},
%the focus is on extracting the distributions on the number of occurrences of $R$.
%Although this could be useful in our case as well (e.g., for frequent pattern mining),
For the task of forecasting, a useful distribution that can be calculated is the so-called
waiting-time distribution. 
The waiting-time for a pattern $R$ when its \dfasr$\ $ is in state $q$ is a random variable,
denoted by $W_{R}(q)$.
It is defined as the number of transitions until its
first full match, i.e., until the DFA visits for the first time one of its final states.
%
%More formally, the waiting-time is the following random variable (the smallest time index at which the DFA visits a final state):
%
%\begin{equation*}
%W_{R}=inf\{n: Y_{0},Y_{1},...,Y_{n}, Y_{n} \in F\}
%\end{equation*}
%
%The above definition assumes that we start from the first symbol in the stream and that the initial state is the DFA's start state.
%A definition more useful for online forecasting would be:
%More formally:
%\begin{definition}[Waiting-time]
\begin{equation*}
W_{R}(q)=inf\{n: Y_{0},Y_{1},...,Y_{n}, Y_{0}=q, q \in Q \backslash F, Y_{n} \in F\}
\end{equation*}
%\end{definition}
The DFA is in a non-final state $q$ and we are interested in the smallest time index $n>0$ (i.e., first time) at which it will visit a final state. 
Informally, what we want to achieve through $W_{R}(q)$ is the following:
each time the DFA is in some non-final state $q$ (regardless of whether it is the start state),
we want to estimate how many transitions we will have to wait until it reaches one of
its final states, i.e., until a full match is detected. 
This number of future transitions may be given to the user as a forecast and it is
constantly revised as more symbols are consumed and the DFA moves to other states.
As a random variable, $W_{R}(q)$ follows a probability distribution and our aim
is to compute this distribution for every non-final state $q$.

%Since we are interested only in the first visit to a final state 
%(in general, using the theory behind Markov chains, 
%we can compute the distributions for the second, third, etc. visits),
We can compute the distribution of $W_{R}(q)$ through the following technique.
First, we convert each state of \pmcmr$\ $ that corresponds to 
a final state $f$ of \dfasr$\ $ ($f \in F$, with $\lvert F \rvert{=}k$) into an absorbing state,
i.e., a ``sink'' state with probability of staying in the same state equal to $1.0$ (state 3 in Figure \ref{fig:dfa_mc_example}).
We can then re-organize the transition matrix as follows:
\begin{equation}
\label{eq:matrix}
\boldsymbol{\Pi} = 
\begin{pmatrix} 
\boldsymbol{N} & \boldsymbol{C}  \\ 
\boldsymbol{0} & \boldsymbol{I}
\end{pmatrix}
\end{equation}
where $\boldsymbol{I}$ is the identity matrix of size $k \times k$, 
corresponding to the absorbing states.
If \pmcmr$\ $ has a total of $l$ states ($k$ of which are final),
then $\boldsymbol{N}$ would be of size $ (l-k) \times (l-k)$ and would correspond to the non-final states, 
holding the probabilities for all the possible transitions between (and only between) the
non-final states. 
Finally, $\boldsymbol{C}$ is a $(l-k) \times k$ matrix holding the transition probabilities
from non-final to final states and $\boldsymbol{0}$ is a zero matrix of size $k \times (l-k)$. 
For example, for the PMC of Figure \ref{fig:mctcc0}, the transition matrix would be the following:
\begin{equation*}
\label{eq:matrix_example}
\boldsymbol{\Pi} = 
\begin{Bmatrix} 
0 \\ 1 \\ 2 \\ 3 
\end{Bmatrix}
\begin{pmatrix} 
P(b)+P(c) 	& P(a) 		& 0 		& 0 \\
P(b) 		& P(a)		& P(c)		& 0 \\
P(b)		& P(a)		& 0			& P(c) \\
0			& 0			& 0			& 1.0
\end{pmatrix}
\end{equation*}
where, to the left of the matrix, for each of its rows, we show the corresponding states (in curly brackets). In this case, $l{=}4$, $k{=}1$ and $\boldsymbol{N}$ is of size $3 \times 3$.
Through this re-arrangement, we can use the following theorem \cite{fu_distribution_2003}:
\begin{theorem}
Given a transition probability matrix $\boldsymbol{\Pi}$ of a homogeneous Markov chain $Y_{t}$ in the form of Eq.~\eqref{eq:matrix}, the probability for the time index n when the system first enters the set of absorbing states can be obtained from
\begin{equation}
\label{eq:wtd}
P(Y_{n} \in A, Y_{n-1} \notin A,...,Y_{1} \notin A \mid \boldsymbol{\xi_{init}}) =
\boldsymbol{\xi}^{T}\boldsymbol{N}^{n-1}(\boldsymbol{I}-\boldsymbol{N})\boldsymbol{1}
\end{equation}
\end{theorem}
$A$ denotes the set of absorbing states.
$\boldsymbol{1}$ is simply a $(l-k) \times 1$ vector with all its elements equal to $1.0$.
$\boldsymbol{\xi_{init}}$ is the initial distribution on the states,
i.e., it is a vector whose element $i$ holds the probability that the PMC
is in state $i$ at the start.
$\boldsymbol{\xi}$ consists of the $l-k$ elements of $\boldsymbol{\xi_{init}}$ corresponding to non-absorbing states.

In the theory of Markov chains,
the current state of the chain is not always known and must be encoded in such a vector.
For example, for the PMC of Figure \ref{fig:mctcc0},
we could have $\boldsymbol{\xi_{init}^{T}}=(0.2\ 0.3\ 0.4\ 0.1)$,
meaning that we are in state 0 with probability $20\%$, in state 1 with probability $30\%$, etc.
However, in our case, at each point, the current state of \dfasr$\ $ 
(and therefore of \pmcmr) is known and therefore this vector would have 
$1.0$ as the value for the element corresponding to the current state (and 0 elsewhere).
$\boldsymbol{\xi}$ changes dynamically as the DFA/PMC moves among its various states and every
state has its own $\boldsymbol{\xi}$, denoted by $\boldsymbol{\xi_{q}}$:
%\begin{equation*}
%\boldsymbol{\xi_{q}}^{T} = (0\ ...\ 1\ ...\ 0)
%\end{equation*}
%where 
\[ \boldsymbol{\xi_{q}}(i) =
  \begin{cases}
    1.0 & \quad \text{if row } i \text{ of } N \text{ corresponds to state } q   \\
    0  & \quad \text{otherwise} \\
  \end{cases}
\]
%For the example of Figure \ref{fig:mctcc0} and for its transition matrix given by Equation \ref{eq:matrix_example}, we have:
%\begin{equation*}
%\boldsymbol{\xi_{0}}^{T} = (1\ 0\ 0)\ \ \boldsymbol{\xi_{1}}^{T} = (0\ 1\ 0)\ \ \boldsymbol{\xi_{2}}^{T} = (0\ 0\ 1)
%\end{equation*}
 
A slight variation of Equation \ref{eq:wtd} then gives the probability of the waiting-time variable:
\begin{equation*}
P(W_{R}(q)=n)=\boldsymbol{\xi_{q}}^{T}\boldsymbol{N}^{n-1}(\boldsymbol{I}-\boldsymbol{N})\boldsymbol{1}
\end{equation*}

\section{Implementation}
\label{sec:implementation}

We implemented a forecasting system, Wayeb, based on Pattern Markov Chains.
Algorithm \ref{alg:wayeb} presents in pseudo-code the steps taken for
recognition and forecasting.
Wayeb reads a given pattern $R$ in the form of a regular expression,
%creates the regular expression $\Sigma^{*}R$ and 
transforms this expression into a NFA
and subsequently, through standard determinization algorithms,
the NFA is transformed into a \textit{m-unambiguous} DFA (line \ref{alg:wayeb:builddfa} in Algorithm \ref{alg:wayeb}).
For the task of event recognition, only this DFA is involved.
At the arrival of each new event (line \ref{alg:wayeb:newevent}),
the engine consults the transition function of the DFA
and updates the current state of the DFA (line \ref{alg:wayeb:updatedfa}).
Note that this function is simply a look-up-table,
providing the next state, given the current state and the type of the new event.
Hence, only a memory operation is required.

%Subsequently, based on this DFA,
%its corresponding PMC is constructed.
%If the random process is assumed to be composed of i.i.d. events,
%then constructing the PMC is straightforward.
%If a higher order $m$ is assumed,
%then the DFA needs to be converted to a \textit{m-unambiguous} equivalent
%DFA and then be embedded in a Markov chain.

\begin{algorithm}[t]
\SetAlgoNoLine
\KwIn{Stream $S$, pattern $R$, order $m$, maximum spread $ms$, forecasting threshold \pfc}
\KwOut{For each event $e\in S$, a forecast $I=(start,end)$}
%\nfasr$\ $ = BuildNFA($R$)\;
%\dfasr$\ $ = DeterminizeNFA(\nfasr ,$m$)\;
\dfasr$\ $ = BuildDFA($R$, $m$)\; \label{alg:wayeb:builddfa}
%($TrainStream$,$TestStream$) = SplitStream($S$,$tp$)\;
\pmcmr$\ $ = WarmUp($S$, \dfasr)\; \label{alg:wayeb:warmup}
$F_{\mathit{table}}$ = BuildForecastsTable(\pmcmr , \pfc, $ms$)\; \label{alg:wayeb:buildflut}
$\mathit{CurrentState}$ = $0$\;
$\mathit{RunningForecasts}$ = $\varnothing$\;
\Repeat{$\mathit{true}$}{ 
	$e$ = RetrieveNextEvent($S$)\; \label{alg:wayeb:newevent}
	$CurrentState$ = UpdateDFA(\dfasr , $e$)\; \label{alg:wayeb:updatedfa}
	\eIf{$CurrenState$ not final} {
		$I$ = $F_{\mathit{table}}(\mathit{CurrentState})$\; \label{alg:wayeb:newforecast}
		$\mathit{RunningForecasts} = I \cup \mathit{RunningForecasts}$ \label{alg:wayeb:storeforecast}
		%add $I$ to $RunningForecasts$\;
	}{
		UpdateStats($\mathit{RunningForecasts}$)\; \label{alg:wayeb:updatestats}
		$\mathit{RunningForecasts}$ = $\varnothing$\; \label{alg:wayeb:clearforecasts}
	}
}
\caption{Wayeb}
\label{alg:wayeb}
\end{algorithm}

\subsection{Learning the matrix of the PMC}
%With respect to forecasting,
To perform event forecasting,
we need to create \pmcmr$\ $ and estimate its transition matrix.
This is achieved by using the maximum-likelihood estimators for the transition
probabilities of the matrix \cite{lothaire_applied_2005}.
Let $\boldsymbol{\Pi}$ denote the transition matrix of a 1-order Markov chain, 
$\pi_{i,j}$ the transition probability from state $i$ to state $j$ 
and $n_{i,j}$ the number of transitions from state $i$ to state $j$. 
Then, the maximum likelihood estimator for $\pi_{i,j}$ is given by:
\begin{equation}
\label{eq:pi_estim}
\hat{\pi}_{i,j}=\frac{n_{i,j}}{\sum_{k \in Q} n_{i,k}}=\frac{n_{i,j}}{n_{i}}
\end{equation}
where $n_{i}$ denotes the number of visits to state $i$. 
Note also that we slightly abuse notation in the above formula,
by using the symbol $Q$, which usually refers to the set of states of the DFA,
to also denote the set of states of the PMC.

In order to obtain a realization of the sequence $Y$ of the states that \dfasr$\ $ visits and the observed values $\hat{\pi}_{i,j}^{obs}$ as estimates for the
transition probabilities,
we can use an initial warm-up period during which a part of the stream
is fed into the engine,
the number of visits and transitions are counted and the transition probabilities are calculated, as per Equation \eqref{eq:pi_estim} (line \ref{alg:wayeb:warmup} in Algorithm \ref{alg:wayeb}). 

\subsection{Building forecasts}
\label{sec:buildforecasts}

After estimating the transition matrix,
\pmcmr$\ $ is used in order to compute the waiting-time distributions
for each non-final state.
Based on these waiting-time distributions,
we build the forecasts associated with each state (line \ref{alg:wayeb:buildflut}).
A forecast produced by Wayeb is in the form of an interval $I=(start,end)$.
The meaning of this interval is the following:
at each point, the DFA is in a certain state.
Given this state, we forecast that the DFA will have reached its final state
(and therefore the pattern fully matched)
at some future point between $start$ and $end$,
with probability at least \pfc.
%Therefore, $start\geq 1$ and $end\geq start$.
The calculation of this interval is done by using the waiting-time distribution that corresponds to each state
and the threshold \pfc\ is set beforehand by the user.

\begin{figure}[!ht]
    \centering
    \begin{subfigure}[b]{0.44\textwidth}
        \includegraphics[width=\textwidth]{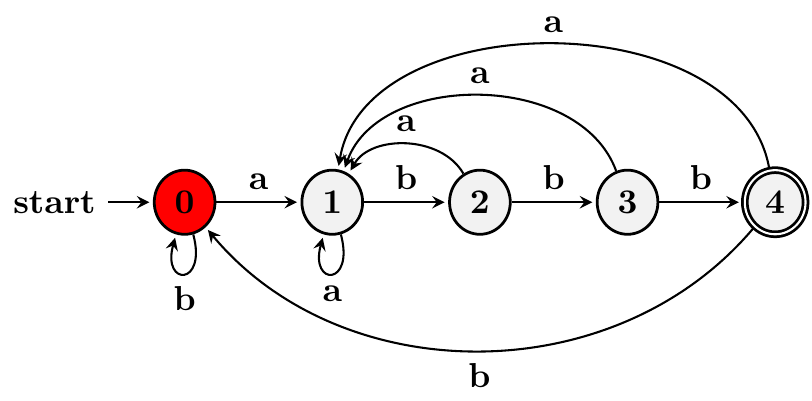}
        \caption{DFA, state 0.}\label{fig:dfa0}
    \end{subfigure}
    \begin{subfigure}[b]{0.44\textwidth}
        \includegraphics[width=\textwidth]{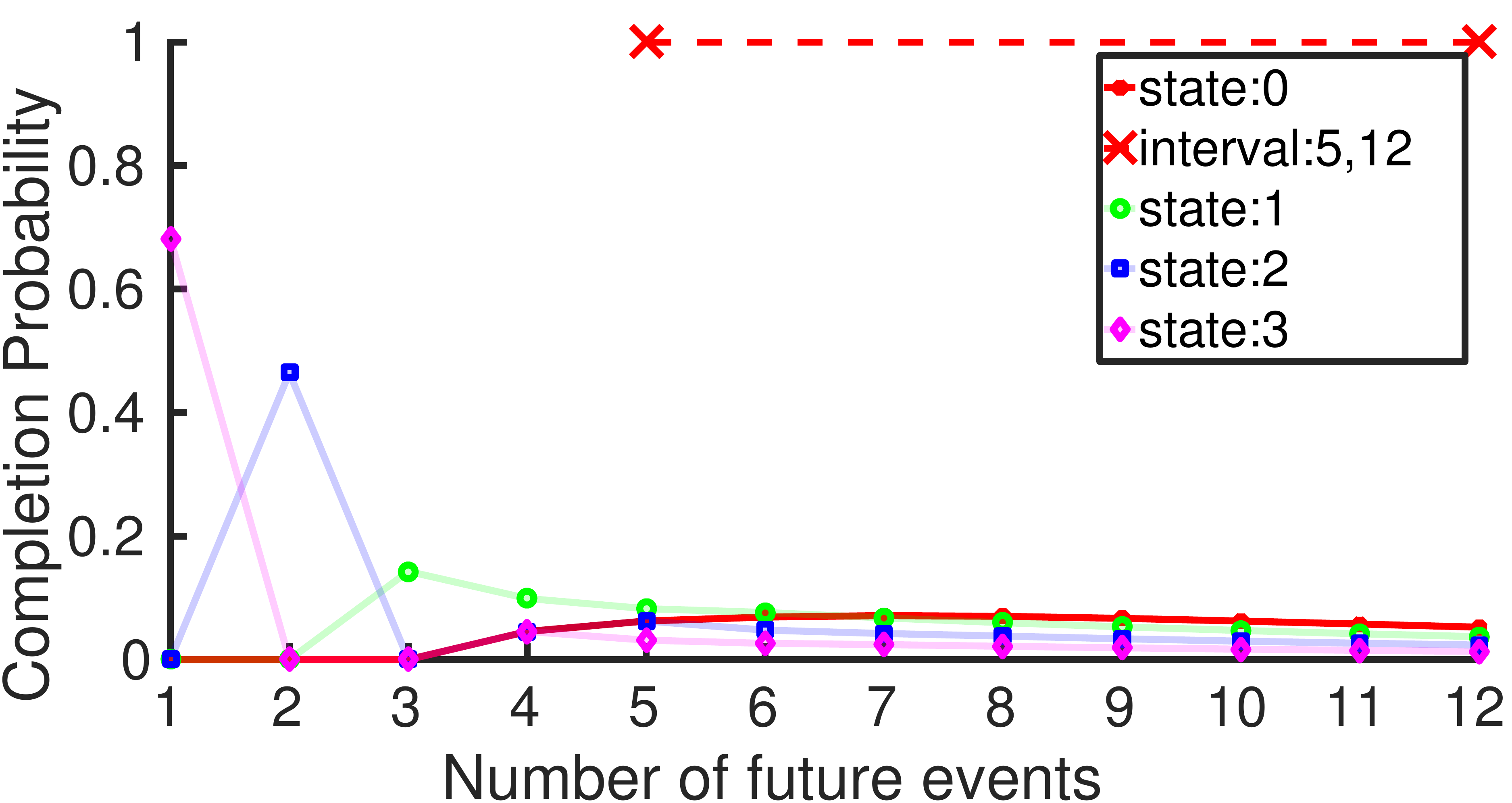}
        \caption{Waiting-time distribution, state 0.}\label{fig:wt0}
    \end{subfigure}
    
    \hfill
    
    \begin{subfigure}[b]{0.44\textwidth}
        \includegraphics[width=\textwidth]{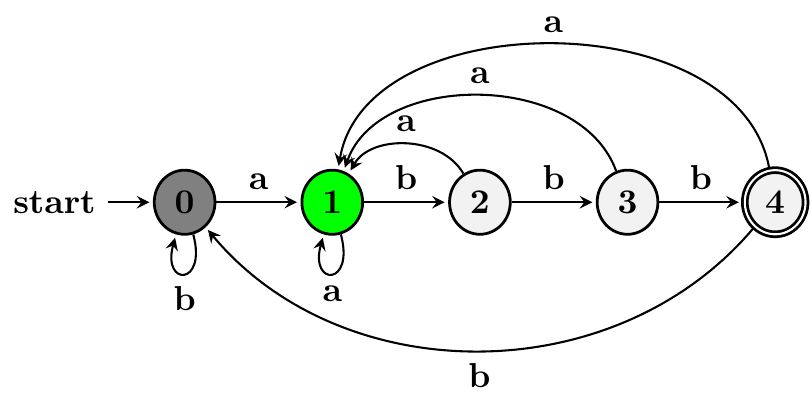}
        \caption{DFA, state 1.}\label{fig:dfa1}
    \end{subfigure}
    \begin{subfigure}[b]{0.44\textwidth}
        \includegraphics[width=\textwidth]{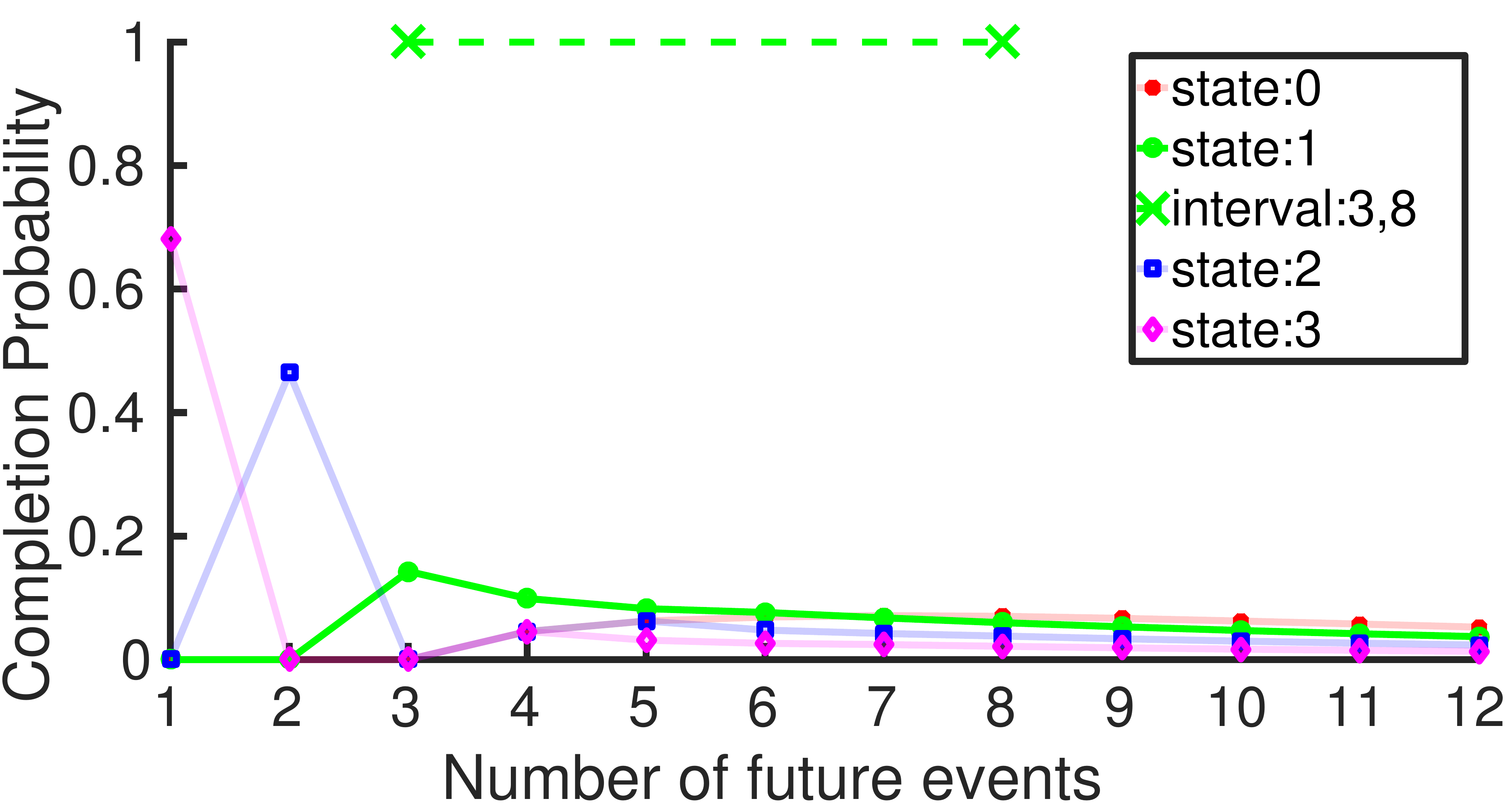}
        \caption{Waiting-time distribution, state 1.}\label{fig:wt1}
    \end{subfigure}
    
    \hfill
    
    \begin{subfigure}[b]{0.44\textwidth}
        \includegraphics[width=\textwidth]{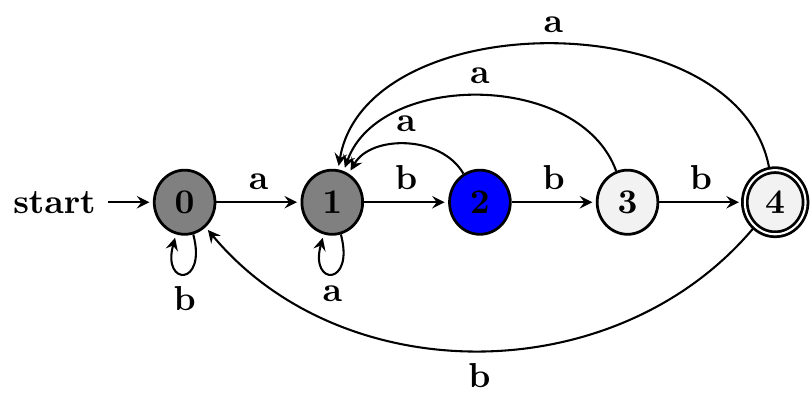}
        \caption{DFA, state 2.}\label{fig:dfa2}
    \end{subfigure}
    \begin{subfigure}[b]{0.44\textwidth}
        \includegraphics[width=\textwidth]{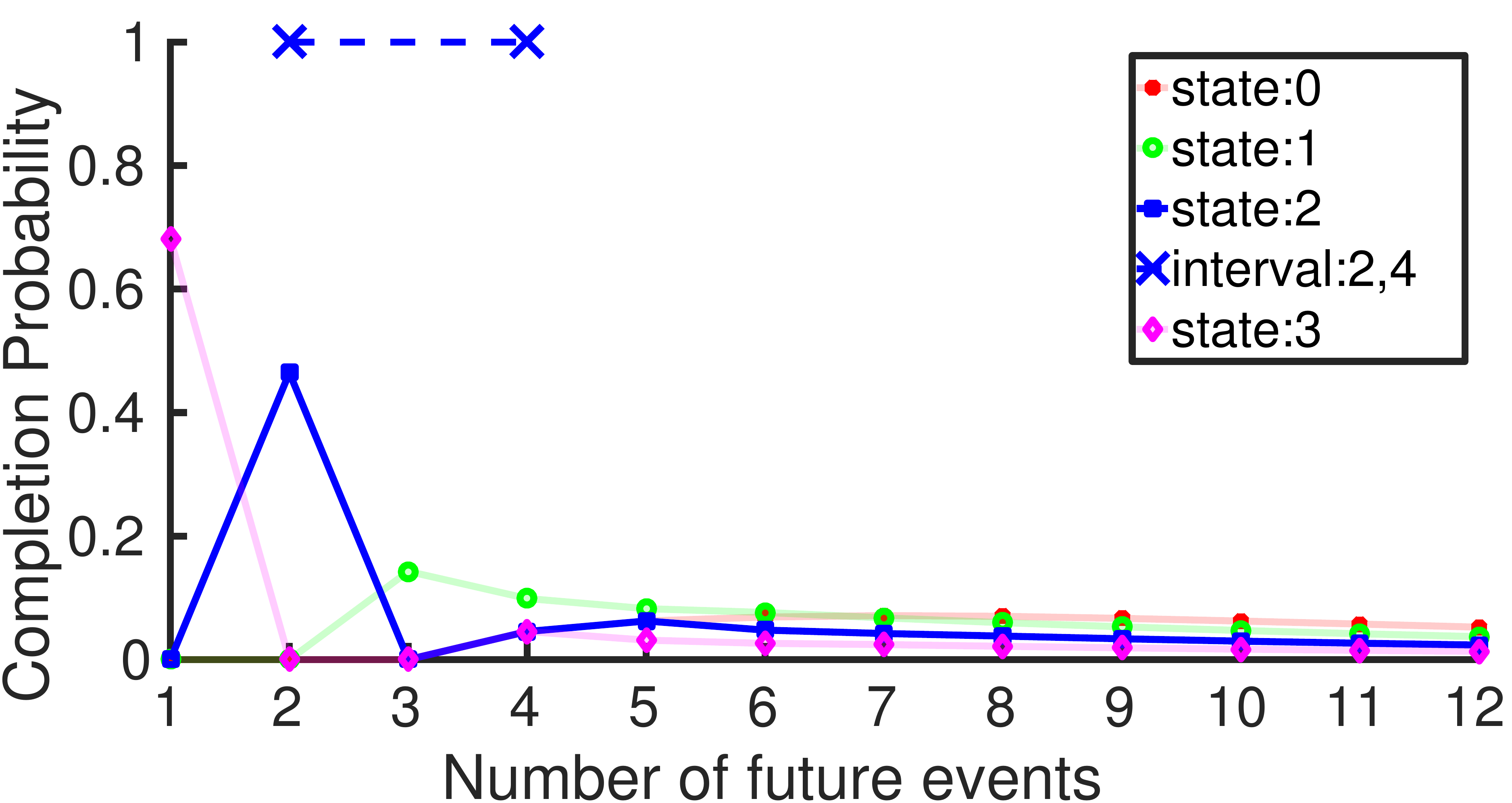}
        \caption{Waiting-time distribution, state 2.}\label{fig:wt2}
    \end{subfigure}
    
    \hfill
    
    \begin{subfigure}[b]{0.44\textwidth}
        \includegraphics[width=\textwidth]{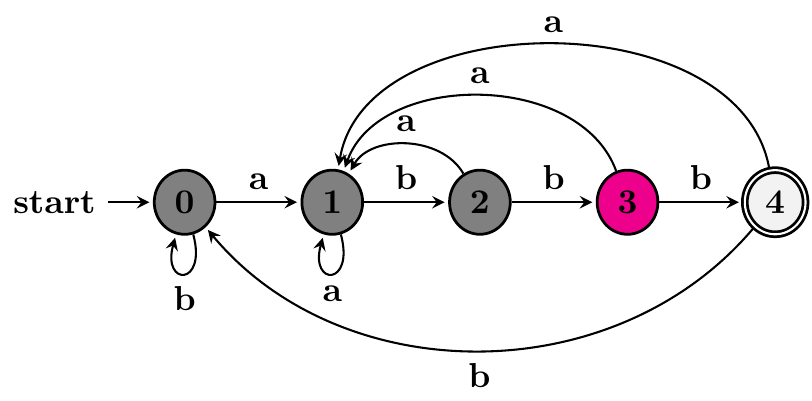}
        \caption{DFA, state 3.}\label{fig:dfa3}
    \end{subfigure}
    \begin{subfigure}[b]{0.44\textwidth}
        \includegraphics[width=\textwidth]{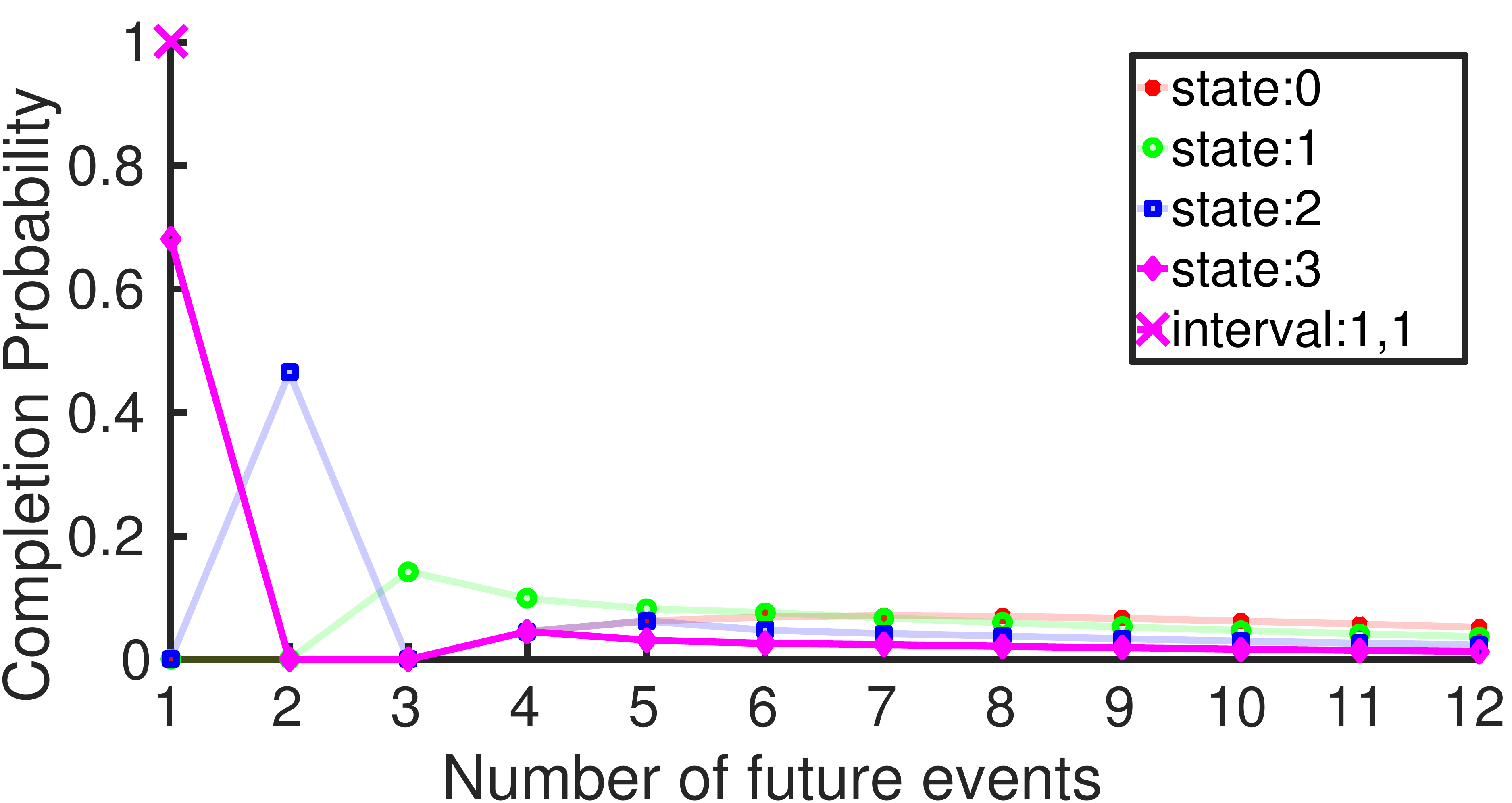}
        \caption{Waiting-time distribution, state 3.}\label{fig:wt3}
    \end{subfigure}

\caption{Example of how forecasts are produced. The pattern $R$ is a sequential pattern $R=a\cdot b\cdot b\cdot b$ (one event of type $a$ followed by three events of type $b$). $\Sigma=\{a,b\}$ (only two event types may be encountered, $a$ and $b$) and $m=1$. No maximum threshold for spread is set. $P_{\mathit{fc}}=0.5$. For illustration purposes, the $x$ axes stop at 12 future events.}\label{fig:wtdfas}
\end{figure}

Each interval $I$ that may be defined on the waiting-time distribution has an associated probability, given by:
\begin{equation*}
P(I)=\sum_{n \in I}{P(W_{R}(q)=n)}
\end{equation*}
where we sum the probabilities of all points $n$ that fall within $I$ ($start\leq n\leq end$, where $n$ is discrete).
We define the set of intervals $I_{\mathit{fc}}$ as:
\begin{equation*}
I_{\mathit{fc}} = \{I: P(I) \geq P_{\mathit{fc}}\}
\end{equation*}
i.e., out of all possible intervals,
$I_{\mathit{fc}}$ contains those that have a probability above the user-defined threshold \pfc.
Any one of the intervals in $I_{\mathit{fc}}$ may be provided as a forecast.
However, a lengthy interval 
(e.g., the whole domain of the distribution has probability $100\%$
and therefore always belongs to $I_{\mathit{fc}}$)
is less informative than a small one.
Therefore, out of all the intervals in $I_{\mathit{fc}}$,
we wish to provide the one that has the smallest length.
We define the spread of an interval as:
\begin{equation*}
spread(I) = end - start
\end{equation*}
The forecast is therefore given as:
\begin{equation}
\label{eq:interval}
I_{\mathit{best}} = \argmin_{I \in I_{\mathit{fc}}} spread(I)
\end{equation}

If more than one interval with the same smallest spread exist,
then we choose the one with the highest probability.
From each waiting-time distribution,
we extract the best interval, as defined by Equation \eqref{eq:interval},
using a single-pass algorithm that scans the distribution of each state only once. 
We may additionally require that the spread of the forecast interval is no greater than a specified maximum threshold $ms$ 
(see relevant input argument in line \ref{alg:wayeb:buildflut} of Algorithm \ref{alg:wayeb}).
In this case, however, it might not be possible to find
an interval that satisfies both constraints
\begin{equation*}
P(I)\geq P_{\mathit{fc}} \wedge spread(I)\leq ms
\end{equation*}
and the algorithm will return an empty interval.

An example of how forecasts are produced is shown in Figure \ref{fig:wtdfas}.
The pattern $R$ is a simple sequential pattern $R=a\cdot b\cdot b\cdot b$ (one event of type $a$ followed by three events of type $b$).
Also $\Sigma=\{a,b\}$ (only two event types may be encountered, $a$ and $b$) and $m=1$. 
Therefore, the distributions are calculated based on the conditional probabilities $P(a{\mid}a)$, $P(a{\mid}b)$, $P(b{\mid}a)$ and $P(b{\mid}b)$.
No maximum threshold for the spread has been set in this example.
As shown in Figure \ref{fig:dfa0}, the DFA has 5 states (0-4) and state 4 is the final state.
When no event has arrived (or only $b$ events have arrived), the DFA is in its start state.
The waiting-time distribution for this state is shown in Figure \ref{fig:wt0} as the red curve.
The other distributions are shown as well, but they are greyed out,
indicating that only the red curve is ``activated'' in this state.
If the user has set $P_{fc}=0.5$,
then the best interval that Wayeb can produce is the one shown above the distributions
(red, dashed line), and this is $I=(5,12)$.
Notice that, as expected,
according to the red distribution, it is impossible that the pattern is fully matched within the next three events
(it is in the start state and needs to see at least 4 events).
If an $a$ event arrives, the DFA moves to its next state, state 1 (Figure \ref{fig:dfa1}),
and now another distribution is ``activated'' (green curve, Figure \ref{fig:wt1}). 
The best interval is now $I=(3,8)$ and has a smaller spread.
The arrival of a $b$ event activates the blue distribution (Figure \ref{fig:wt2}) and this time an even smaller interval is produced, $I=(2,4)$.
If a second $b$ event arrives, the magenta distribution is activated.
This distribution has a peak above $0.5$ which is the value of the threshold \pfc\
and this allows the engine to produce an interval with a single point $I=(1,1)$.
Essentially, Wayeb informs us that, with probability at least $50\%$,
we will see a full match of the pattern in exactly 1 event from now.

Note that the calculation of the forecast intervals for each state
needs to be performed only once,
since for the same state it results always in the same interval being computed (assuming stationarity, as stated in Section \ref{sec:theory}).
Therefore, the online forecasting system is again composed of a simple look-up-table ($F_{\mathit{table}}$ in line \ref{alg:wayeb:buildflut} of Algorithm \ref{alg:wayeb}) and only memory operations 
%and a simple ``time-shift'' 
are required.
%The time-shift is needed because the originally computed values for $start$ and $end$ have the current state as a reference point, whatever that might be.
%For example, if we have seen 100 events and the forecast interval for the current state is $(2,5)$,
%then the final interval produced is $(102,105)$.

\subsection{Performance and quality metrics}

There are three metrics that we report in order to assess Wayeb's performance and the quality of its forecasts:
\begin{itemize}
\item $\mathit{Precision = \frac{\#\ of\ correct\ forecasts}{\#\ of\ forecasts}}$. At every new event arrival, the new state of the DFA is estimated (line \ref{alg:wayeb:updatedfa} of Algorithm \ref{alg:wayeb}).
If the new state is not a final state, a new forecast is retrieved from the look-up-table of forecasts (line \ref{alg:wayeb:newforecast}). These forecasts are maintained in memory (line \ref{alg:wayeb:storeforecast}) until a full match is detected. Once a full match is detected, we can estimate which of the previously produced forecasts are satisfied, in the sense that the full match happened within the interval of a forecast (line \ref{alg:wayeb:updatestats}). These are the correct forecasts. All forecasts are cleared from memory after a full match (line \ref{alg:wayeb:clearforecasts}).
%\item \textit{Recall}: $Recall = \frac{\#\ of\ pattern\ occurrences\ forecast}{\#\ of\ pattern\ occurrences}$. The denominator is provided directly by the recognition module. The numerator counts the number of occurrences for which at least one forecast was satisfied. 
%This counter is updated at every pattern occurrence, by checking the in-memory forecasts.
\item $\mathit{Spread=end - start}$, as described in Section \ref{sec:buildforecasts}.
\item $\mathit{Distance=start - now}$.
This metric captures the distance between the time the forecast is made ($now$) and the earliest expected completion time of the pattern.
Note that two intervals might have the same spread (e.g., $(2,2)$ and $(5,5)$ both have $\mathit{Spread}$ equal to $0$) but different distances ($2$ and $5$, assuming $now=0$).
\end{itemize} 

$\mathit{Precision}$ should be as high as possible.
With respect to $\mathit{Spread}$, the intuition is that, the smaller it is, the more informative the interval.
For example, in the extreme case where the interval is a single point,
the engine can pinpoint the exact number of events that it will have to wait until a full match.
On the other hand, the greater the $\mathit{Distance}$, the earlier a forecast is produced and therefore a wider margin for action is provided.
%By taking all these three metrics into account,
Thus, ``good'' forecasts are those with high precision (ideally $1.0$), 
low spread (ideally $0$) and a distance that is as high as possible 
(ideal values depend on the pattern).
%Note that 
These metrics may be calculated either as aggregates, 
gathering results from all states (in which case average values for $\mathit{Spread}$ and $\mathit{Distance}$ over all states are reported), 
or on a per-state basis, i.e.,
we can estimate the $\mathit{Precision}$, $\mathit{Spread}$ and $\mathit{Distance}$ of the forecasts produced only by a specific state of the DFA.
We omit results for $\mathit{Recall}$ (defined as percentage of detected events correctly predicted by at least one forecast), because $\mathit{Recall}$ values are usually very high and not informative. 

\begin{figure}[t]
    \centering
    \begin{subfigure}[b]{0.45\textwidth}
        \includegraphics[width=\textwidth]{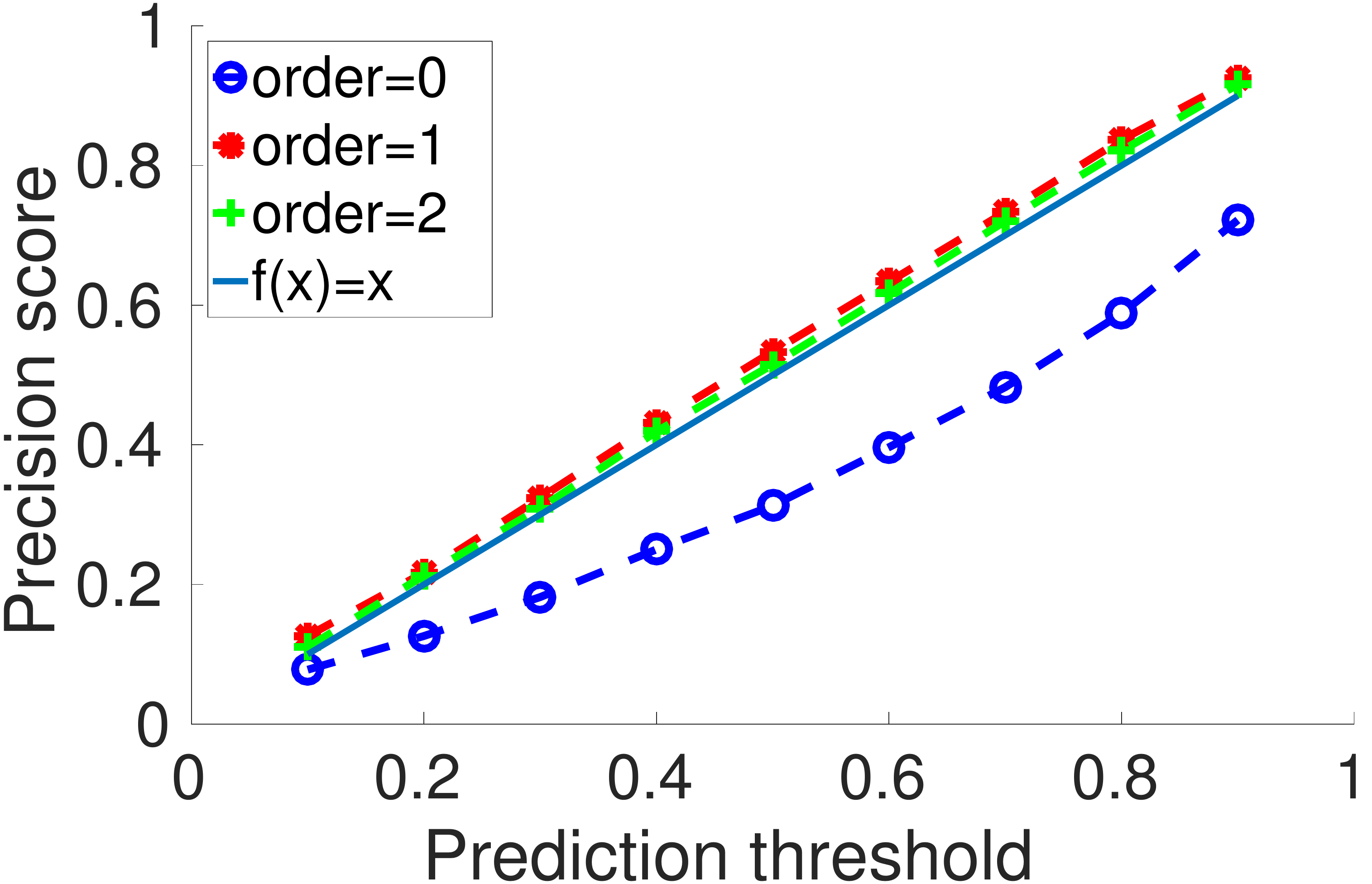}
        \caption{$R = a \cdot b \cdot c$.}\label{fig:valid_abc_prec}
    \end{subfigure}
    \hfill
    \begin{subfigure}[b]{0.45\textwidth}
        \includegraphics[width=\textwidth]{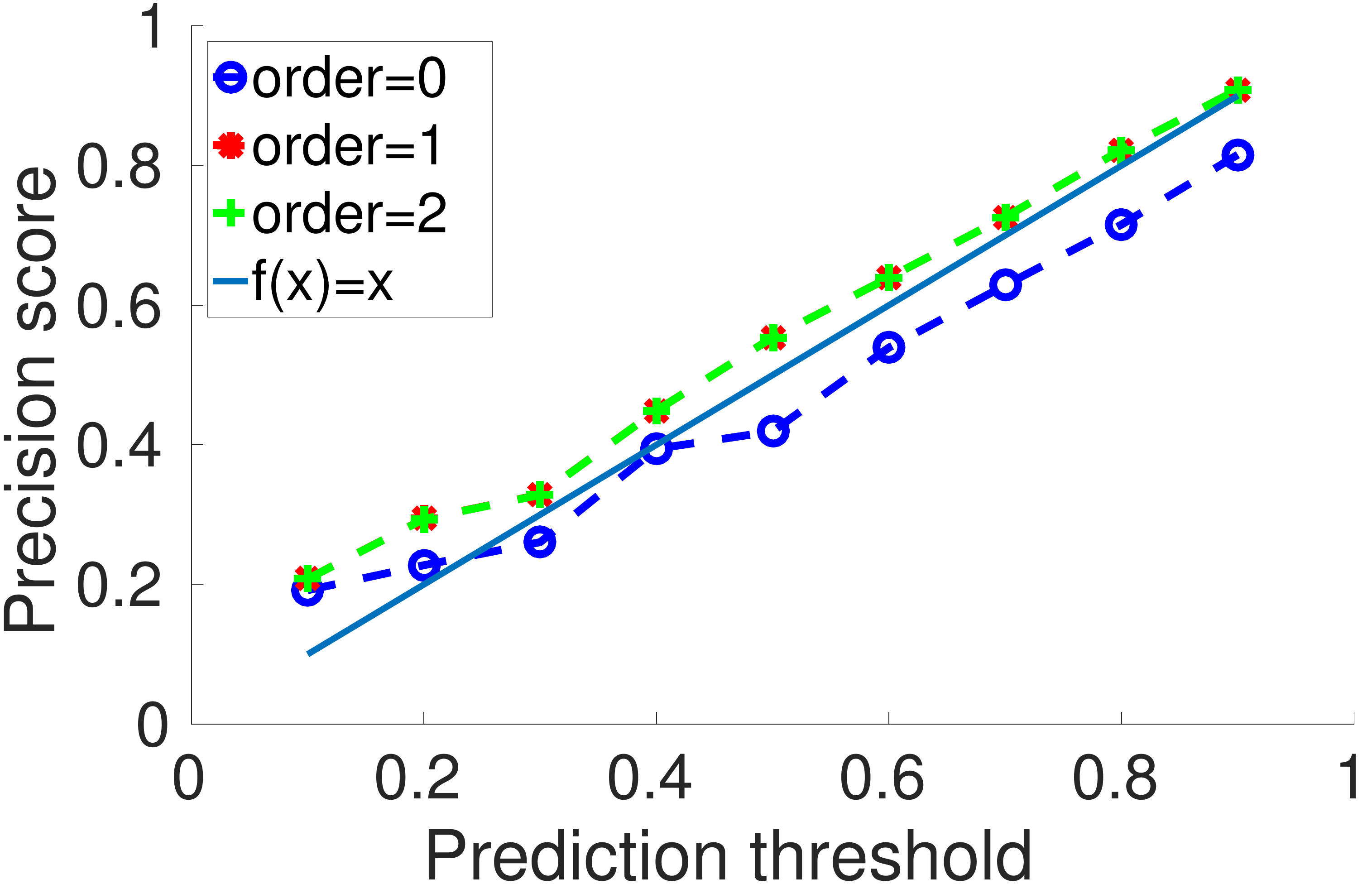}
        \caption{$R = a \cdot (a + b)^{*} \cdot c$.}\label{fig:valid_a_aorbstar_c_prec}
    \end{subfigure}
    \caption{Precision scores for synthetic data, produced by a 1-order Markov process, with $\Sigma=\{a,b,c\}$.}\label{fig:valid_prec}
\end{figure}

\subsection{Validation tests with synthetic data}
\label{sec:experiments:validation}

%In this section we present the results from a series of tests on synthetic datasets.
%Each dataset is provided to Wayeb as a stream of events.
%Part of the stream (usually $\approx 20\%$) is used in order to learn the transition matrix $\Pi$. 
%In each run, only a single pattern is tested and 
%the order $m$, the forecasting threshold \pfc\ and the maximum spread allowed $ms$ are given by the user.

A set of tests was conducted with synthetically generated data for validation purposes. 
Streams were generated by a known Markov process and subsequently the engine was tested on these streams, for various patterns, forecast thresholds and orders. 
Figure \ref{fig:valid_prec} shows the aggregate (from all states) precision scores for two patterns, tested against a stream produced by a 1-order Markov process.
The first pattern is the simple sequence $R = a \cdot b \cdot c$.
The second pattern, $R = a \cdot (a + b)^{*} \cdot c$, is more complex and involves a \textit{star closure} operation on the \textit{union} of $a$ and $b$, right after an $a$ event and before a $c$ event.
For each pattern, three different values of the order $m$ of the PMC were used (0, 1 and 2). 
The figures show how the engine behaves when the forecast threshold is increased and the order $m$ of the PMC changes. 

Note that the line $f(x)=x$ is also included in the figures,
which acts as the baseline performance of the engine.
If the Markov chain constructed for the pattern under test is indeed correct,
then the precision score should lie above this line (or very close).
As described in Section \ref{sec:buildforecasts},
each interval has a probability on the waiting-time distribution of at least $P_{fc}=x$.
Therefore, if these waiting-time distributions are indeed correct,
a percentage of at least \pfc\ of the intervals will be satisfied.
This also means that the actual precision score might be significantly higher than the threshold
in cases where the waiting-time distributions have high peaks.
For example,
in Figure \ref{fig:wt3} the single-point interval produced has a probability of $\approx 70\%$,
hence $\approx 70\%$ of forecasts from that state will be satisfied,
which is significantly higher than the $50\%$ forecasting threshold.

For both of the tested patterns, 
when $m=0$ (blue curves) the precision scores are below the baseline performance,
indicating that a PMC without memory is unable to produce satisfactory forecasts
for a 1-order stream.
When $m$ is increased to match the order of the generating process ($m=1$, red curves), the precision score does indeed lie above the baseline. 
A further increase in the value of $m$ does not seem to affect the precision score
(in Figure \ref{fig:valid_a_aorbstar_c_prec}, the red and green curves for $m=1$ and $m=2$ respectively coincide completely and only the latter is visible).

In some cases, even if we use an incorrect order $m$,
the precision score may be above the baseline 
or even above the line of the correct order $m$.
This may happen because incorrect models may produce ``pessimistic'' intervals, with high spread and therefore implicitly take a bigger ``chunk'' out of the correct distributions. 
%(for a more detailed explanation, see Appendix \ref{sec:prec_order}).  
In practice, however, 
the spread is constrained for informative forecasts, 
and thus models with incorrect order are insufficient.

\section{Experimental results}
\label{sec:experiments}
We present results from experiments on real-world datasets from credit card fraud management and maritime monitoring.

\subsection{Credit card fraud management}

\begin{figure}[!ht]
    \centering
    \begin{subfigure}[b]{0.32\textwidth}
        \includegraphics[width=\textwidth]{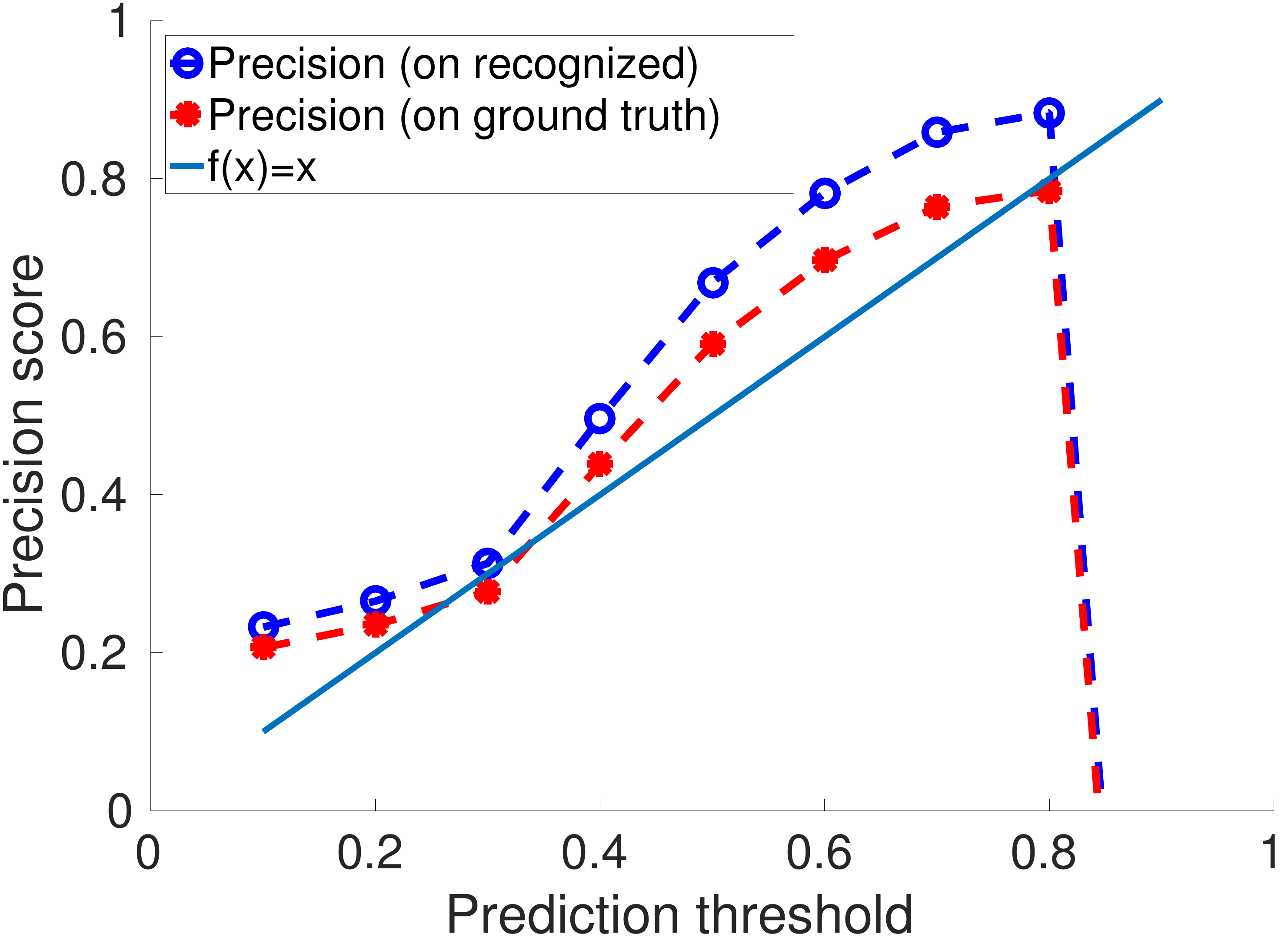}
        \caption{Precision (all states), $m=1$.}\label{fig:real_inc_1_precall}
    \end{subfigure}
    \begin{subfigure}[b]{0.32\textwidth}
        \includegraphics[width=\textwidth]{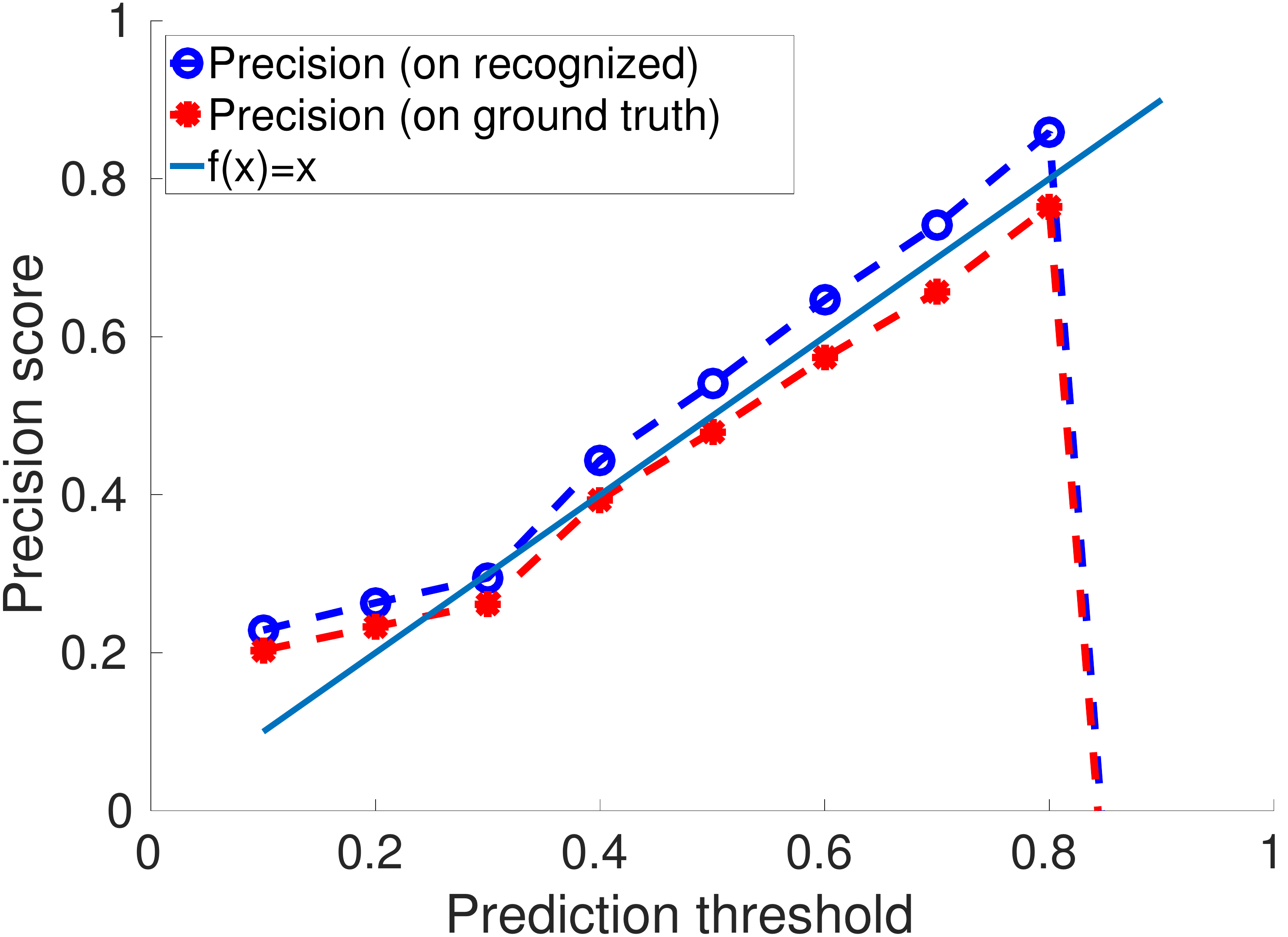}
        \caption{Precision (all states), $m21$.}\label{fig:real_inc_2_precall}
    \end{subfigure}
    \begin{subfigure}[b]{0.32\textwidth}
        \includegraphics[width=\textwidth]{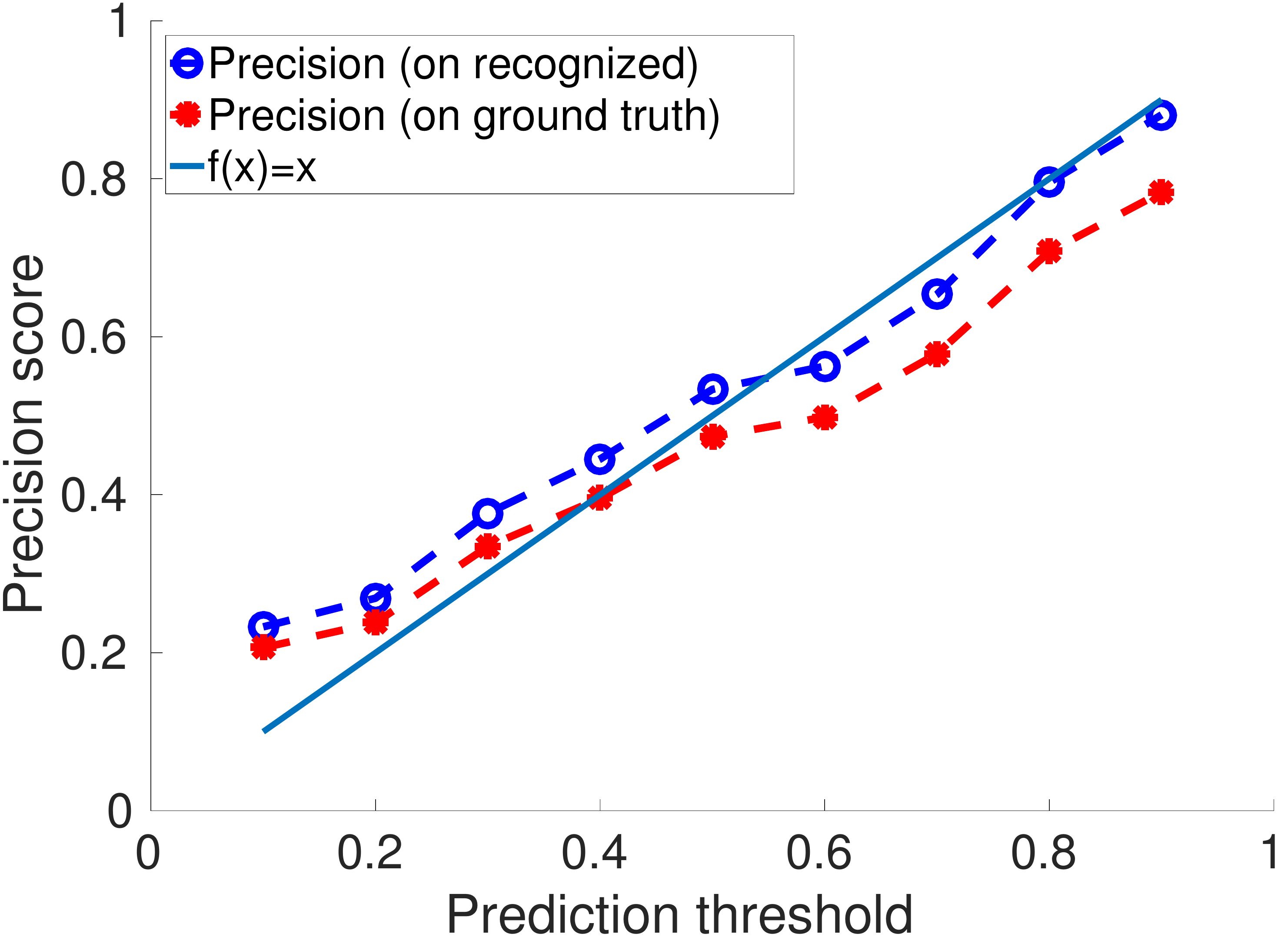}
        \caption{Precision (all states), $m=3$.}\label{fig:real_inc_3_precall}
    \end{subfigure}
    
    \hfill
    
    \begin{subfigure}[b]{0.32\textwidth}
        \includegraphics[width=\textwidth]{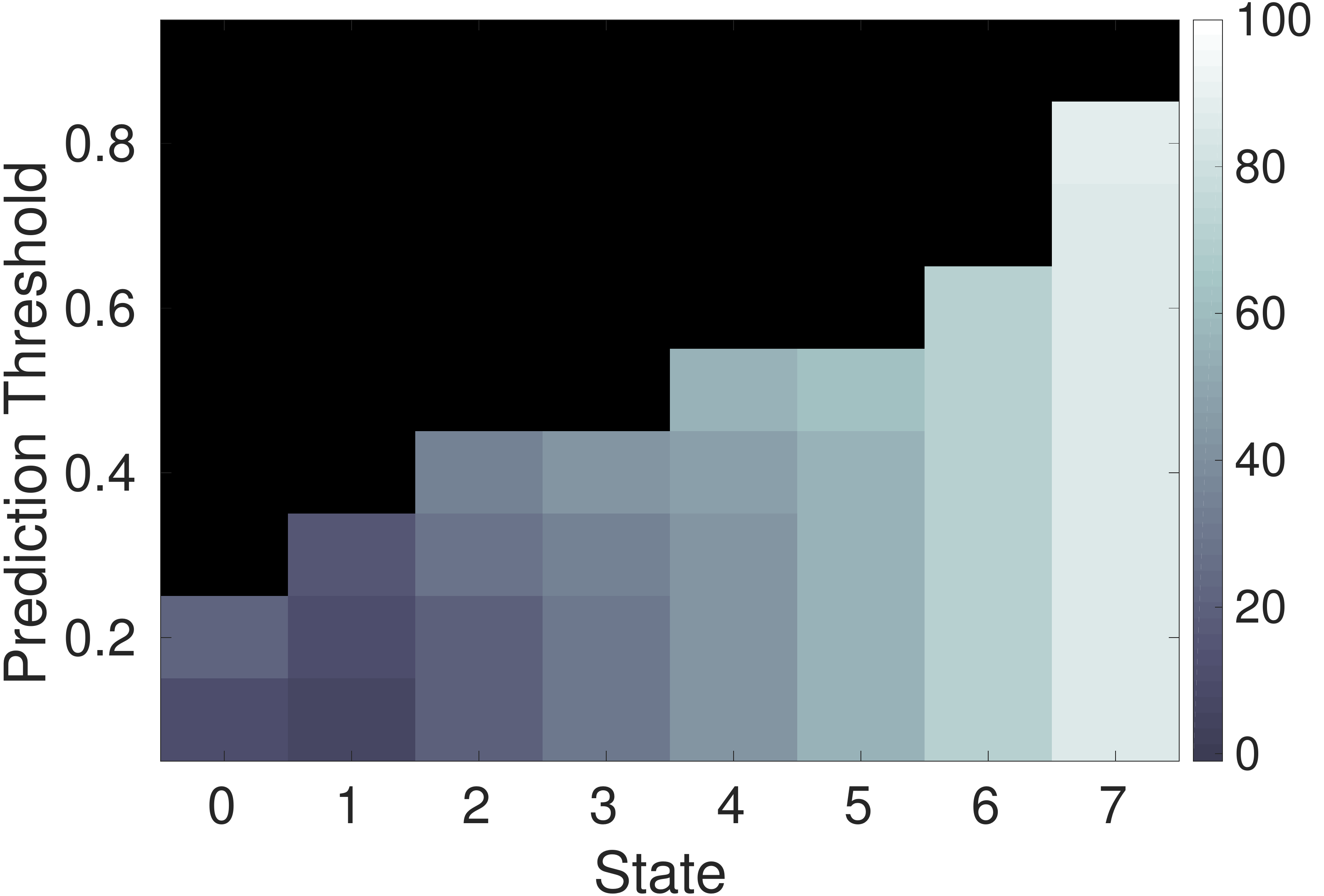}
        \caption{Precision (per state), $m=1$.}\label{fig:real_inc_1_prec}
    \end{subfigure}
    \begin{subfigure}[b]{0.32\textwidth}
        \includegraphics[width=\textwidth]{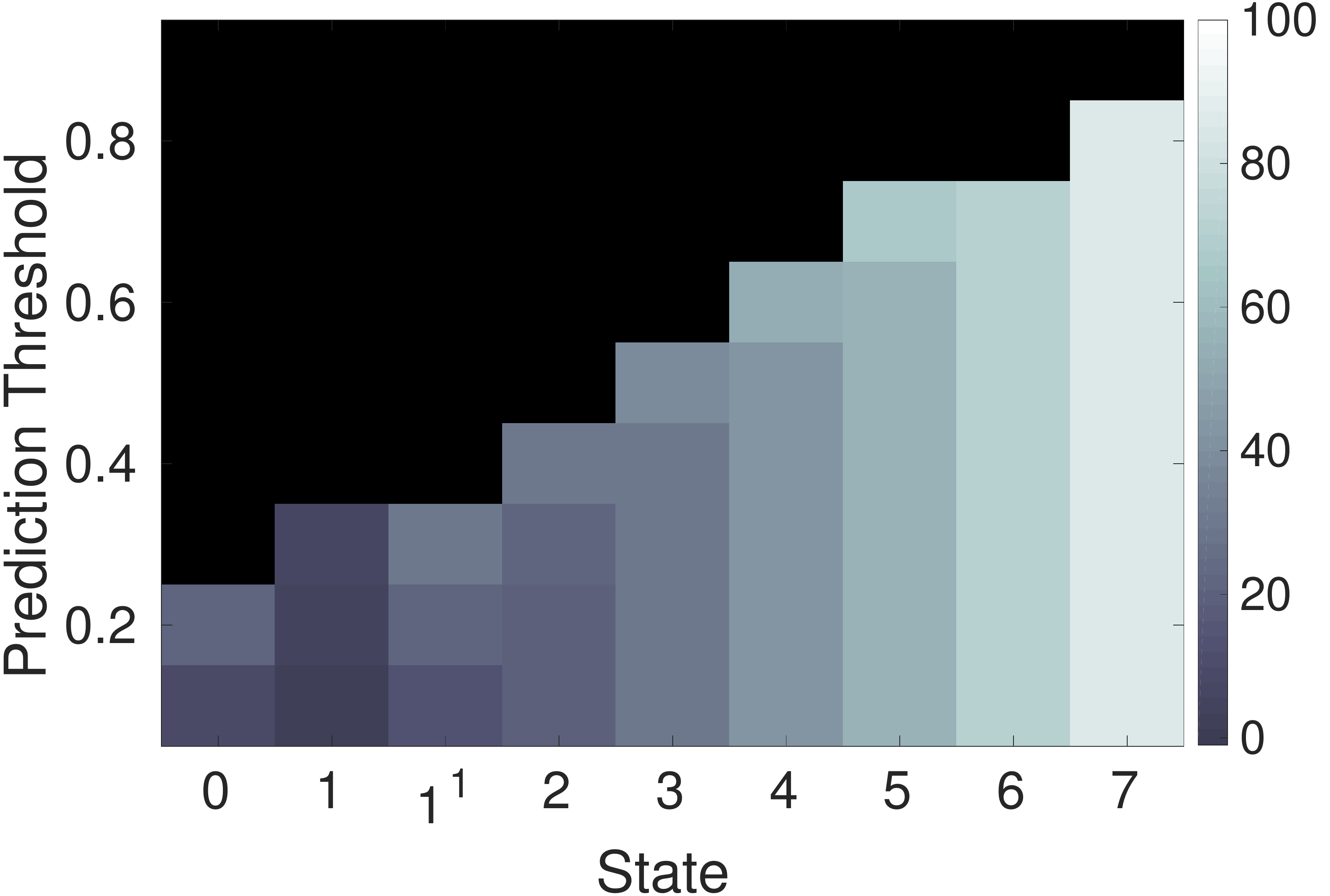}
        \caption{Precision (per state), $m=2$.}\label{fig:real_inc_2_prec}
    \end{subfigure}
    \begin{subfigure}[b]{0.32\textwidth}
        \includegraphics[width=\textwidth]{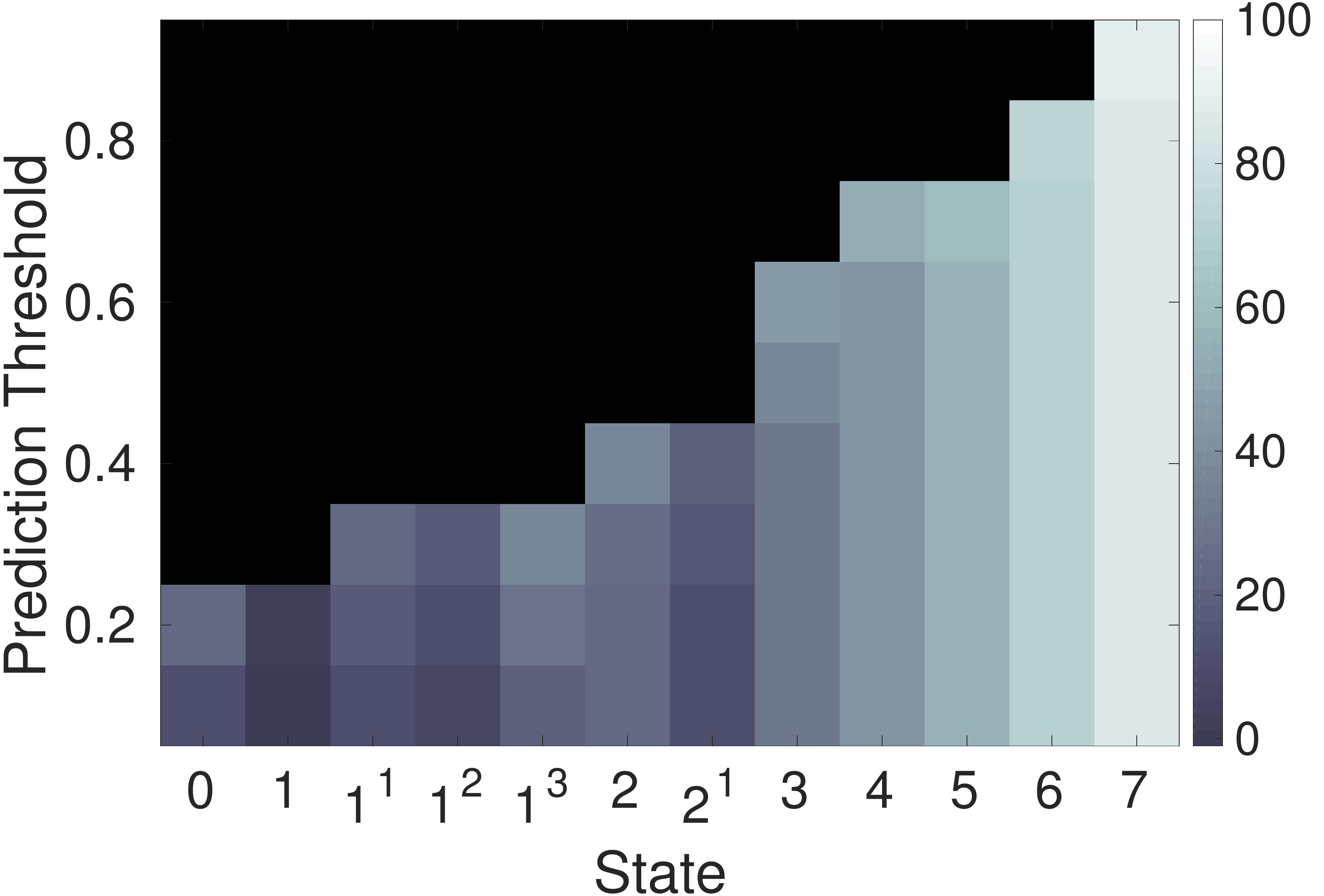}
        \caption{Precision (per state), $m=3$.}\label{fig:real_inc_3_prec}
    \end{subfigure}
    
    \hfill
    
    \begin{subfigure}[b]{0.32\textwidth}
        \includegraphics[width=\textwidth]{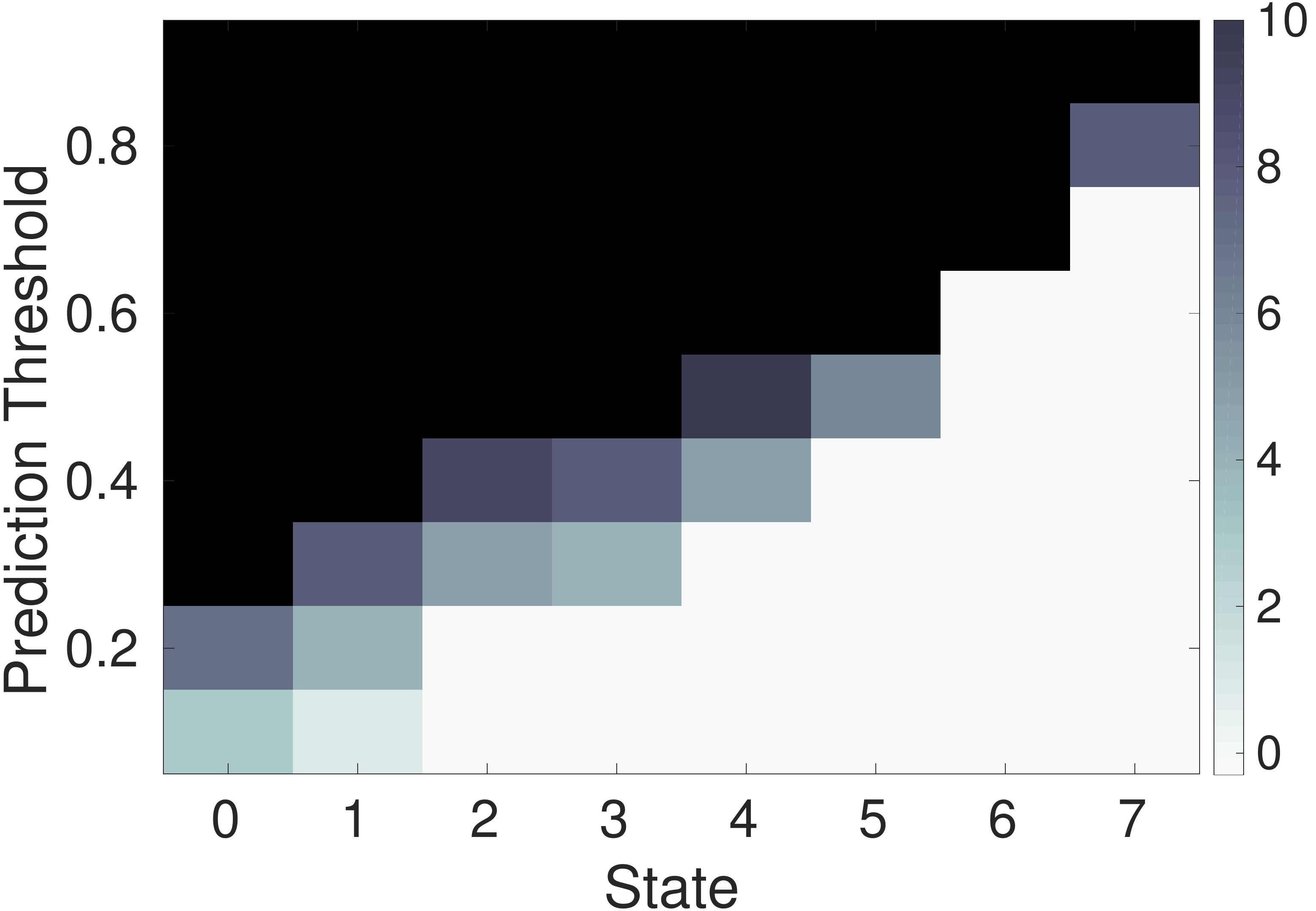}
        \caption{Spread (per state), $m=1$.}\label{fig:real_inc_1_spread}
    \end{subfigure}
    \begin{subfigure}[b]{0.32\textwidth}
        \includegraphics[width=\textwidth]{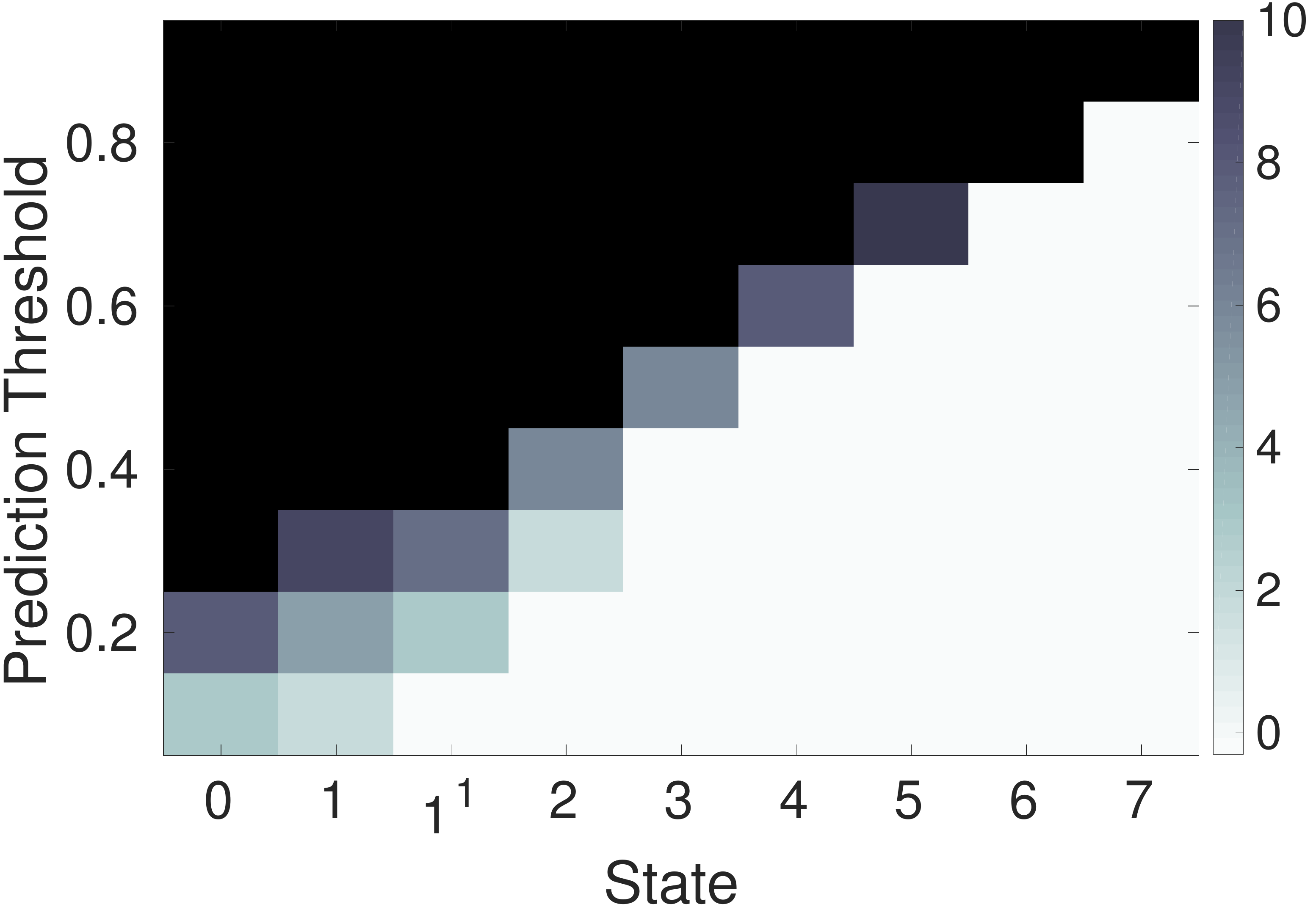}
        \caption{Spread (per state), $m=2$.}\label{fig:real_inc_2_spread}
    \end{subfigure}
    \begin{subfigure}[b]{0.32\textwidth}
        \includegraphics[width=\textwidth]{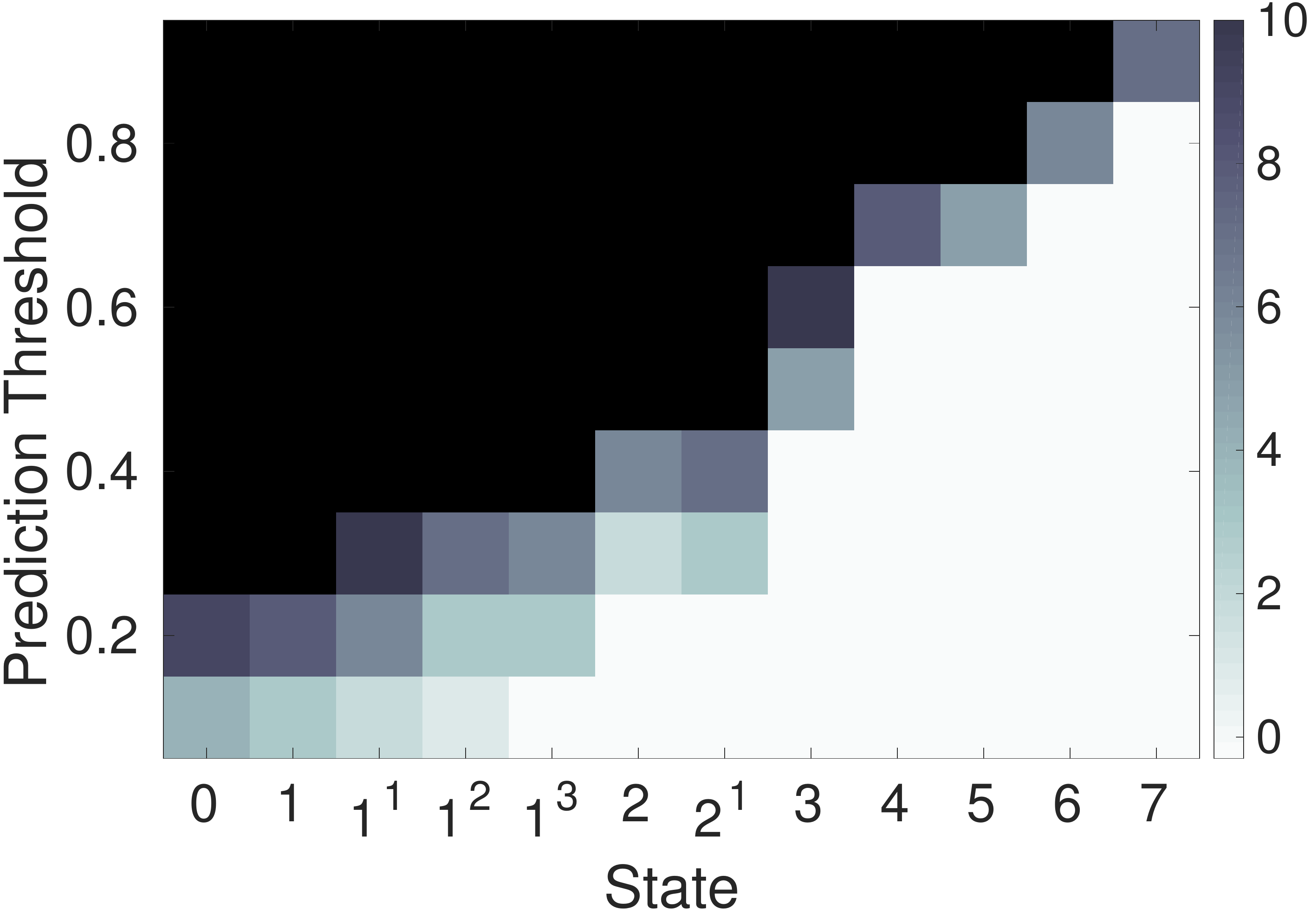}
        \caption{Spread (per state), $m=3$.}\label{fig:real_inc_3_spread}
    \end{subfigure}
    
    \hfill
    
    \begin{subfigure}[b]{0.32\textwidth}
        \includegraphics[width=\textwidth]{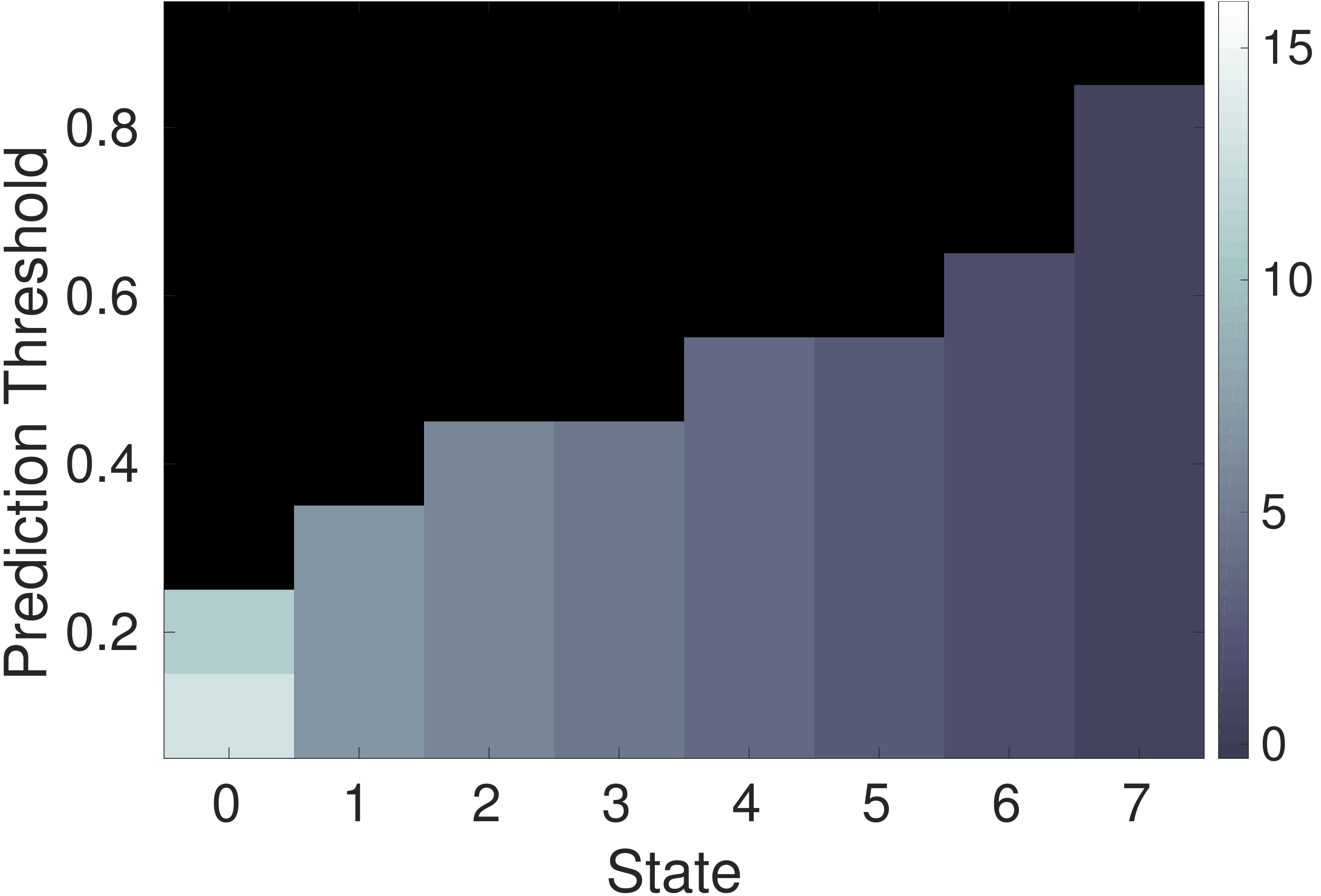}
        \caption{Distance (per state), $m=1$.}\label{fig:real_inc_1_dist}
    \end{subfigure}
    \begin{subfigure}[b]{0.32\textwidth}
        \includegraphics[width=\textwidth]{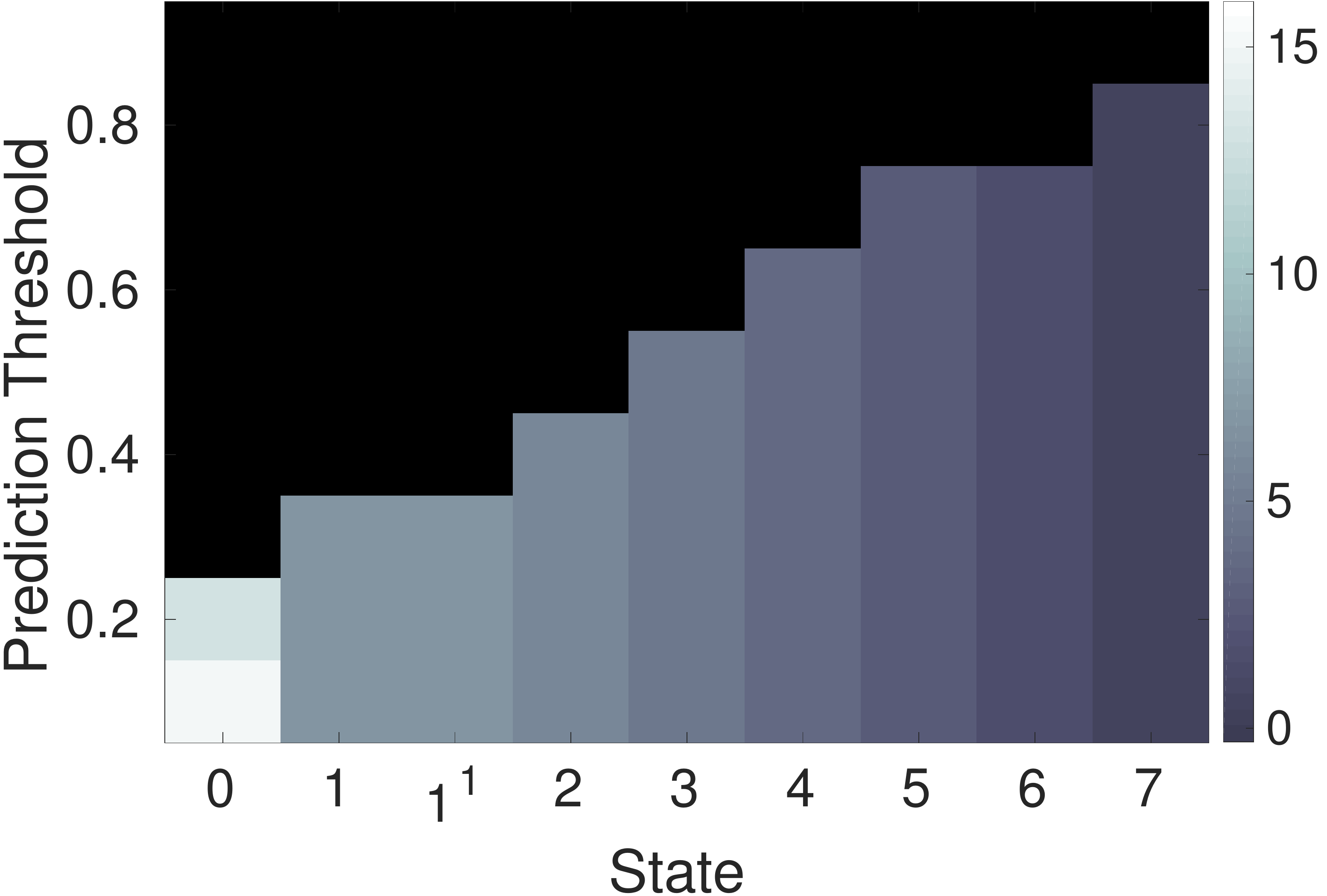}
        \caption{Distance (per state), $m=2$.}\label{fig:real_inc_2_dist}
    \end{subfigure}
    \begin{subfigure}[b]{0.32\textwidth}
        \includegraphics[width=\textwidth]{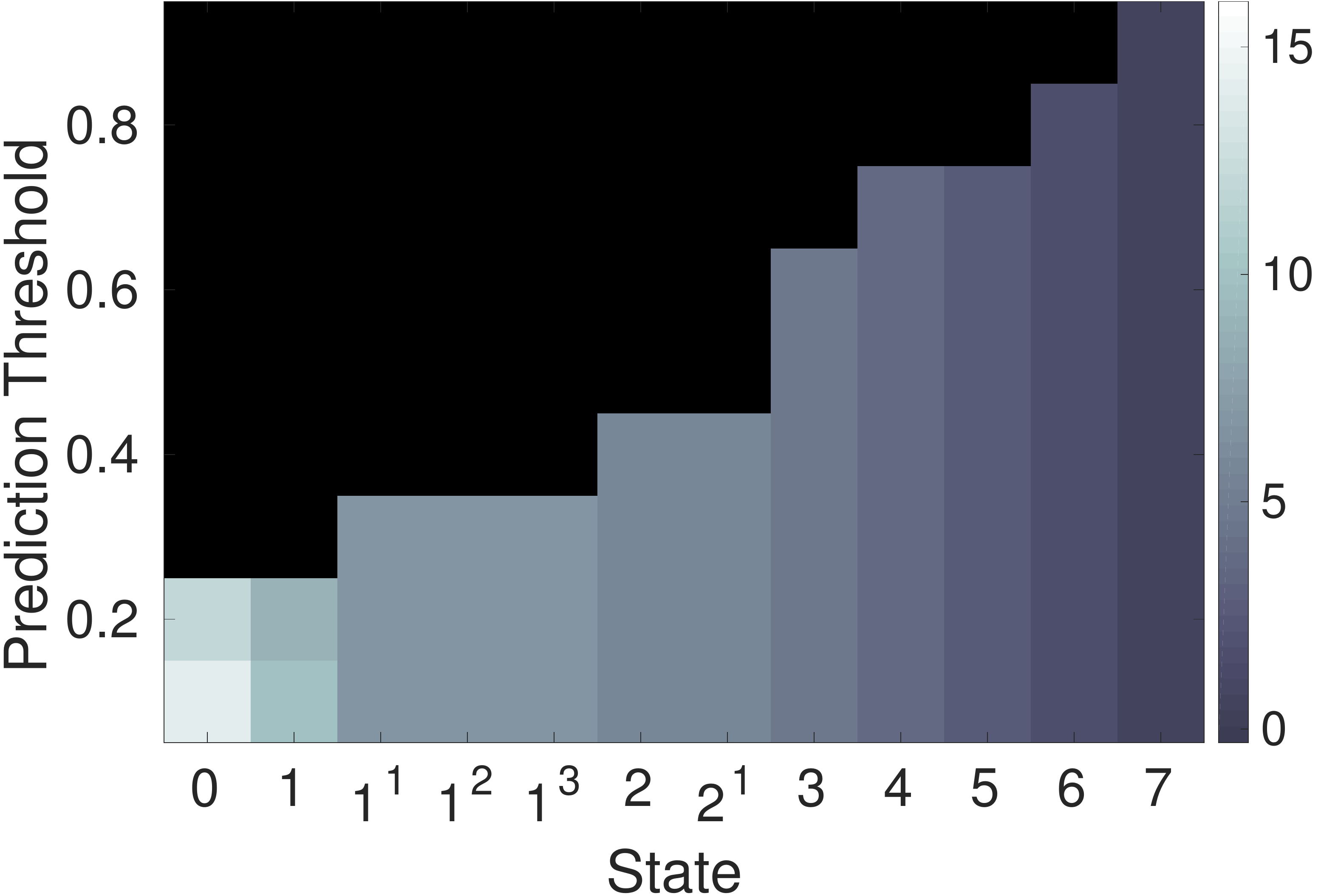}
        \caption{Distance (per state), $m=3$.}\label{fig:real_inc_3_dist}
    \end{subfigure}

\caption{Results for the $\mathit{IncreasingAmounts}$ pattern, for $m=1$, $m=2$ and $m=3$, and for maximum spread $ms=10$.}\label{fig:real_inc_image}
\end{figure}

Unlike most academic and industrial work on fraud management, 
we performed an evaluation on a \emph{real} dataset of credit card transactions, 
made available by 
Feedzai\footnote{\href{https://feedzai.com/}{https://feedzai.com/}}, 
our partner in the 
SPEEDD project\footnote{\href{http://speedd-project.eu/}{http://speedd-project.eu/}}.
Each event is a transaction accompanied by several arguments,
such as the time of the transaction, the card ID, the amount of money spent, etc.
There is also one boolean argument, 
indicating whether the transaction was labeled by a (human) analyst as being fraudulent or not.
The original dataset is highly imbalanced.
Only $\approx 0.2\%$ of the transactions are fraudulent.
We created a summary of this original dataset, 
in which all fraudulent transactions were kept,
but only some of the normal ones,
so that the percentage of fraudulent transactions rises to $\approx 30\%$.
The total number of transactions in this summary dataset was $\approx 1.5$ million.

In order to be able to detect fraudulent transactions,
companies use domain expert knowledge and machine learning techniques,
so that they can extract a set of patterns, indicative of fraud.
For our experiments,
we used a set of fraud patterns provided by Feedzai, our partner in
the SPEEDD project.
We also employed the \textit{parition-contiguity} selection strategy,
where the ID of a card is used as the partition attribute. 
Upon the arrival of a new transaction event, the ID is checked and the event is pushed to the PMC run that is responsible for this ID 
or a new run is created, in case this transaction is the first one for this card.

In Figure \ref{fig:real_inc_image},
the results for the pattern \textit{IncreasingAmounts} are presented,
for three different values of the order $m$ (1, 2 and 3),
where we have set the maximum allowed spread at the value of 10.
This pattern detects 8 consecutive transactions of a card in which the amount of money
in a transaction is higher than the amount in the immediately previous transaction (for the same card ID),
i.e., it attempts to detect sequences of transactions with increasing trends in their amounts. 
Since such direct relational constraints are not currently supported by our engine,
a pre-processing step was necessary. 
During this step, each transaction is flagged as either being \textit{Normal} or 
as one having an \textit{IncreasingAmount} with respect to the immediately previous.
Therefore, the pattern provided to Wayeb starts with one \textit{Normal} transaction,
followed by 7 transactions flagged as \textit{IncreasingAmount}.

Since, in this dataset, there is ground truth available by fraud analysts, 
indicating whether a transaction was fraudulent or not, besides measuring precision with respect to the events detected by the PMC, we can also measure precision with respect to those fraud instances that were both detected and were actually marked as fraudulent. 
The red curves in the precision figures correspond to precision scores as measured when ground truth is taken into account.
Note, however, that the dataset annotation does not contain information about the fraud type. This means that, when we detect a match of the \textit{IncreasingAmounts} pattern and the ground truth informs us that the involved transactions are indeed fraudulent,
there is no way to determine whether they are considered as fraudulent due to a trend of increasing amounts or
to some other pattern. % (e.g., too many transactions within a short time period).
As a result, the red curves could be ``optimistic''. 

For all three values of the order $m$, Wayeb can maintain a precision score that lies above the
$f(x){=}x$ line (Figures \ref{fig:real_inc_1_precall} and \ref{fig:real_inc_2_precall}) or is very close to it (Figure \ref{fig:real_inc_3_precall}), i.e., the produced forecasts, compared against the recognized matches (blue curves), satisfy the threshold constraint.
However, when $m{=}1$ or $m{=}2$ and $P_{fc}{=}0.9$,
Wayeb cannot find intervals whose probability is at the same time above this threshold and whose spread is below 10, and fails
to produce any forecasts (the sudden drop in the curves indicates forecast unavailability).
By increasing the order to $m{=}3$ and taking more past events into account (Figure \ref{fig:real_inc_3_precall}),
Wayeb can handle this high forecast threshold
(we will come back to this issue at the end of this section).
As compared against the ground truth (red curves),
the precision scores are lower.
This precision discrepancy between scores estimated against recognized matches and scores estimated against ground truth
is due to the fact that the fraud pattern is imperfect,
i.e., there are cases with 8 consecutive transactions with \textit{IncreasingAmount} which do not actually constitute fraud. 
It is interesting to note, though, that the shape of the red curves closely follows that of the blue ones,
indicating that, by using a more accurate pattern, 
we would indeed be able to achieve ground truth precision closer to that of the blue curves for all values of \pfc.

The precision scores of Figures \ref{fig:real_inc_1_precall}, \ref{fig:real_inc_2_precall} and \ref{fig:real_inc_3_precall} are calculated by combining the forecasts produced by all states of the PMC. 
In order to better understand Wayeb's behavior,
a look at the behavior of individual states could be more useful.
Figures \ref{fig:real_inc_1_prec} -- \ref{fig:real_inc_3_dist} depict image plots for various metrics against both the forecast threshold and the state of the PMC.
The metrics shown are those of precision (on the recognized matches), spread and distance.
We omit the plots for ground truth precision because they have the same shape as those for precision on recognized matches, but with lower values.
In each such image plot the $y$ axis corresponds to the various values of \pfc.
The $x$ axis corresponds to the states of the PMC.
Each state has a unique integer identifier, starting from $0$ (the start state).
We group together states that are duplicates of each other,
in cases where some states are ambiguous.
For example, in Figure \ref{fig:real_inc_3_prec},
states $1^{1}$, $1^{2}$ and $1^{3}$ are all duplicates of state 1.
In this way, the $x$ axis shows how advanced we are in the recognition process,
when moving from one cluster of duplicates to the next.
The black areas in these plots are ``dead zones'',
meaning that, for the corresponding combinations of \pfc\ and state,
Wayeb fails to produce forecasts (i.e., it cannot guarantee, according to the learned transition probabilities, that the forecast intervals will have at least \pfc\ probability of being satisfied).
On the contrary, areas with light colors are ``optimal'',
in the sense that they have high precision,
low spread (the colorbar is inverted in the spread plots)
and high distance in their respective plots.

The precision plots (\ref{fig:real_inc_1_prec}, \ref{fig:real_inc_2_prec}, \ref{fig:real_inc_3_prec}) show that the more advanced states of the PMC enter into such dead zones at higher forecast thresholds. 
Figures \ref{fig:real_inc_1_spread}, \ref{fig:real_inc_2_spread} and \ref{fig:real_inc_3_spread} show the spread of the forecast intervals.
Two clearly demarcated zones emerge.
One is the usual dead zone (black, top left).
The other one (white, bottom right) corresponds to forecasts whose spread is 0,
i.e., single point forecasts. 
A common behavior for all states is that, as higher values of \pfc\ are set to the engine,
the spread increases, i.e., each state attempts to satisfy the constraint of the forecast threshold
by taking a longer interval from the waiting-time distribution.
On the other hand, those states that are more advanced can maintain small spread values for a
wider margin of \pfc\ values. 
For example, in Figure \ref{fig:real_inc_1_spread}, 
state 2 maintains a spread of $0$ until $P_{fc}{=}0.2$ whereas state 5 hits this
limit at around $P_{fc}{=}0.5$.
Figures \ref{fig:real_inc_1_dist}, \ref{fig:real_inc_2_dist} and \ref{fig:real_inc_3_dist}
show image plots for the distance metric.
As can be seen,
those regions that have high precision scores and low spread,
also tend to have low distance.
Therefore, there is a trade-off between these three metrics.
For the case of $m{=}1$,
good forecasts might be considered those of state 5,
which can maintain a small spread until $P_{fc}{=}0.5$ 
and whose temporal distance is $\approx 3$.
By increasing $m$,
one can get good forecasts that are more ``satisfactory'',
at the cost of an increased size for the PMC
(a discussion about this cost will be presented shortly).
For example, when $m{=}2$,
as in Figure \ref{fig:real_inc_2_spread},
state 5 can produce single point forecasts for higher values of \pfc\ (for $0.6$ too).

As a final comment, we note that increasing the value of $m$ does not necessarily imply higher precision scores.
In fact, as shown in Figures \ref{fig:real_inc_1_precall}, \ref{fig:real_inc_2_precall}, \ref{fig:real_inc_3_precall},
the precision score might even decrease.
This behavior is due to the fact that, in general,
smaller values of $m$ tend to produce more ``pessimistic'' intervals,
with higher spread.
For example, for $P_{fc}{=}0.8$, the precision score for $m{=}1$ (Figure \ref{fig:real_inc_1_precall}) is in fact higher than when $m{=}2$ (Figure \ref{fig:real_inc_2_precall}).
In Figure \ref{fig:real_inc_1_spread},
we can see that, for $m{=}1$, the intervals of state 7 when $P_{fc}{=}0.8$ (the only state producing forecasts for this value of \pfc) have a high spread whereas the same state, when $m{=}2$, produces intervals with low spread (Figure \ref{fig:real_inc_2_spread}).  
Since ``pessimistic'', high-spread intervals take a bigger ``chunk'' out of a distribution,
their precision scores end up being also higher.
By increasing $m$, Wayeb can approximate the real waiting-time distributions more closely
and therefore produce forecasts with lower spread that are closer to the specified threshold.
Therefore, the accuracy curve (blue, dashed curve) starts to coincide with the $f(x){=}x$ line.

\subsection{Maritime monitoring}

Another real-world dataset against which Wayeb was tested came from the field of maritime monitoring.
When sailing at sea, (most) vessels emit messages relaying information about their position, heading, speed, etc.: the so-called AIS (automatic identification system) messages. 
AIS messages may be processed in order to produce a compressed trajectory, 
consisting of critical points,
i.e., important points that are only a summary of the initial trajectory, 
but allow for an accurate reconstruction \cite{patroumpas_online_2016}.
The critical points of interest for our experiments are the following:
\begin{itemize}
\itemsep0em
\item $\mathit{Turn}$: when a vessel executes a turn.
\item $\mathit{GapStart}$: when a vessel turns off its AIS equipment and stops transmitting its position.
\item $\mathit{GapEnd}$: when a vessel turns on its AIS equipment back again (a $\mathit{GapStart}$ must have preceded).
\end{itemize}   
We used a dataset consisting of a stream of such critical points from $\approx 6.500$ vessels, 
covering a 3 month period and spanning the Greek seas.
Each critical point was enriched with information about whether it is headed towards the
northern, eastern, southern or western direction.
For example, each $\mathit{Turn}$ event was converted to one of $\mathit{TurnNorth}$, $\mathit{TurnEast}$, $\mathit{TurnSouth}$ or $\mathit{TurnWest}$ events.
We show results from a single vessel, with $\approx 50.000$ events.

\begin{figure}[t]
    \centering
    \begin{subfigure}[b]{0.49\textwidth}
        \includegraphics[width=0.8\textwidth]{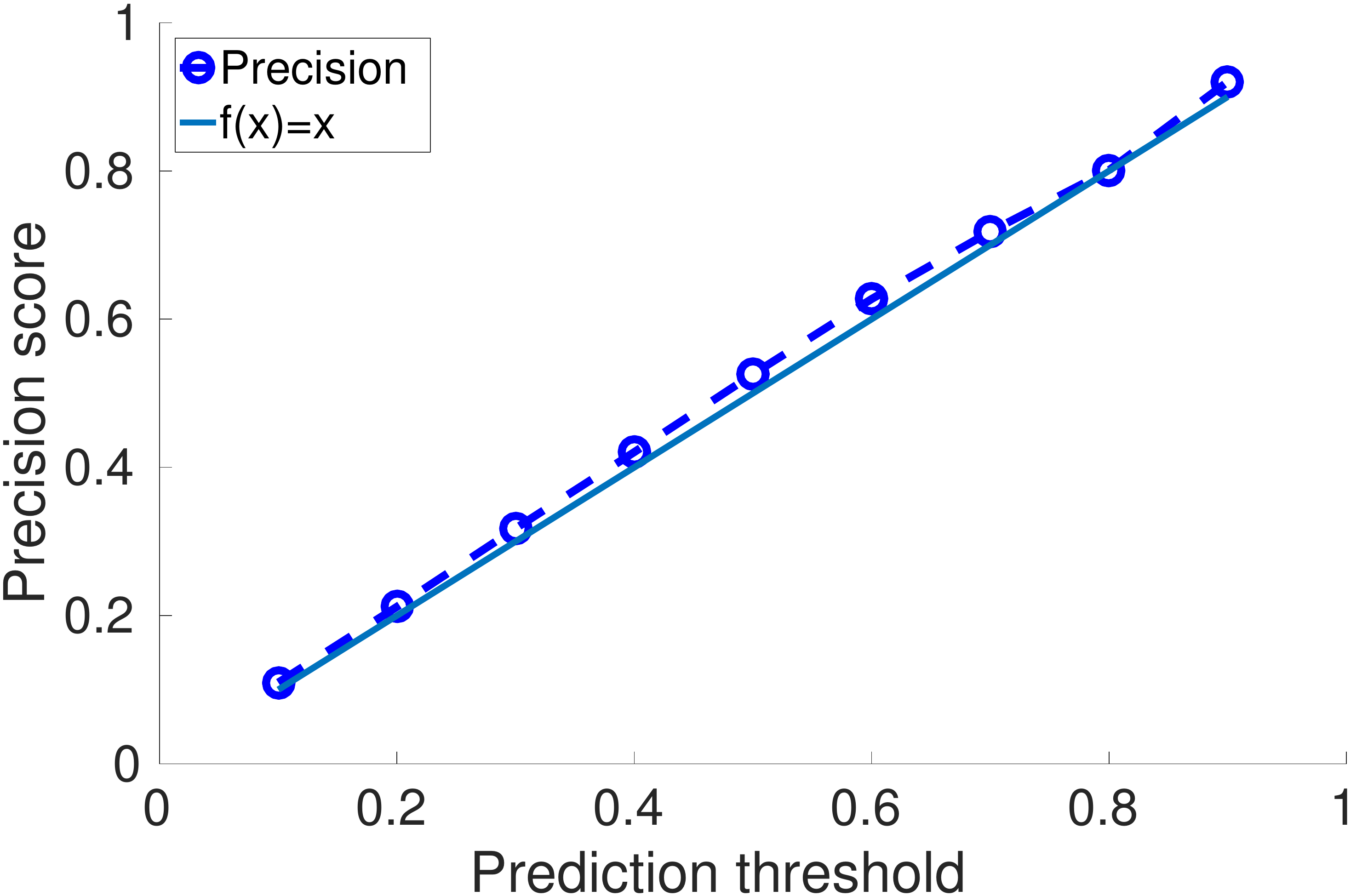}
        \caption{Precision (all states).}\label{fig:real_maritime_1_acc_all}
    \end{subfigure}
    \begin{subfigure}[b]{0.49\textwidth}
        \includegraphics[width=\textwidth]{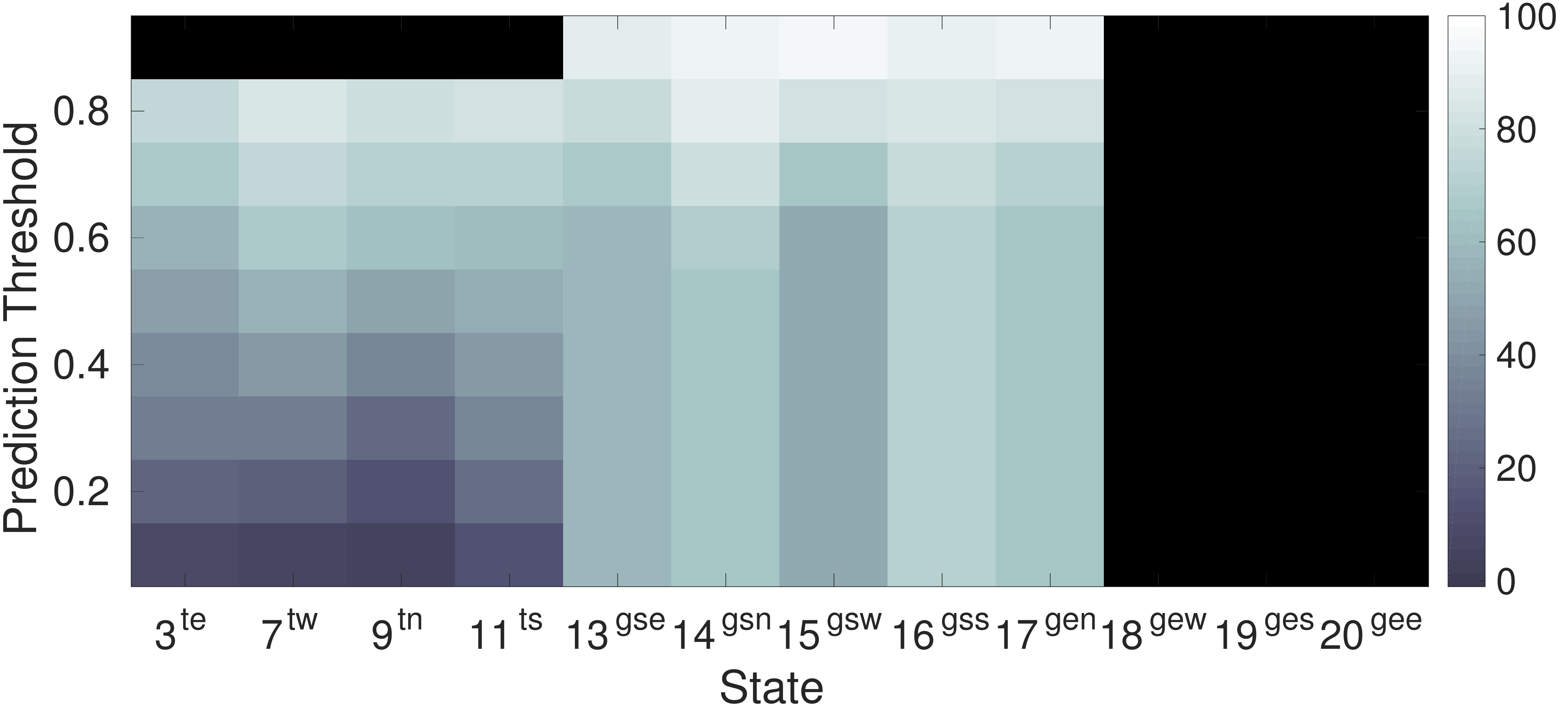}
        \caption{Precision (per state).}\label{fig:real_maritime_1_acc}
    \end{subfigure}
    \hfill
    \begin{subfigure}[b]{0.49\textwidth}
        \includegraphics[width=\textwidth]{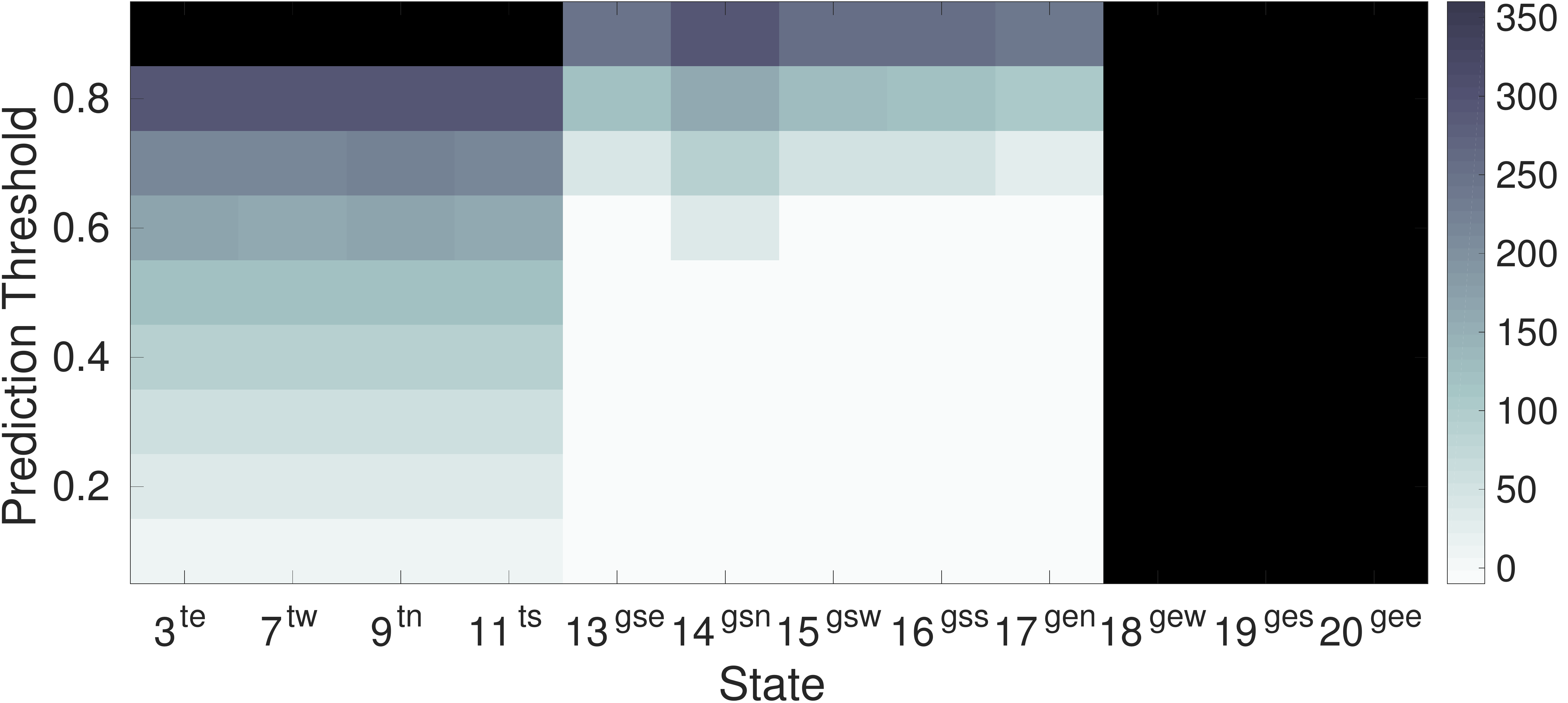}
        \caption{Spread (per state).}\label{fig:real_maritime_1_spread}
    \end{subfigure}
    \begin{subfigure}[b]{0.49\textwidth}
        \includegraphics[width=\textwidth]{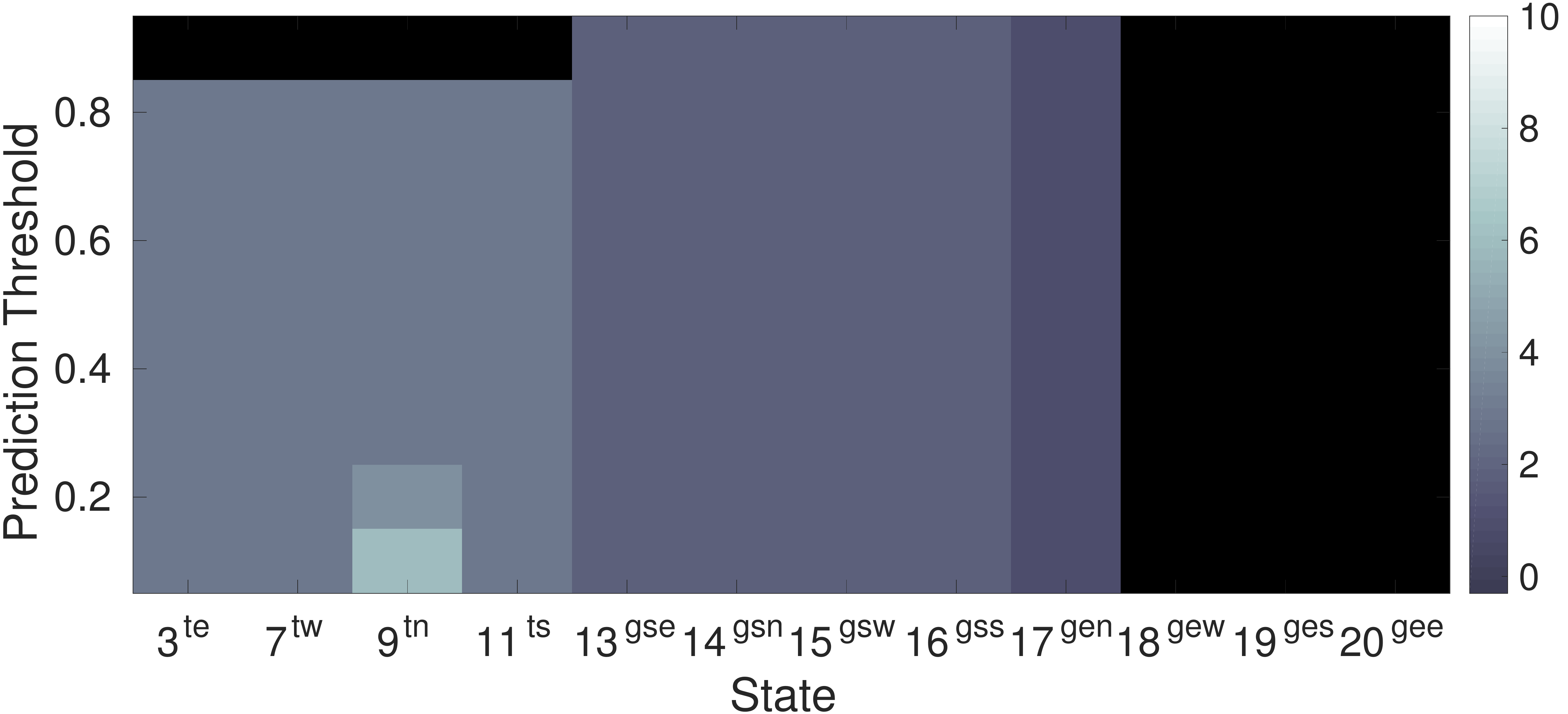}
        \caption{Distance (per state).}\label{fig:real_maritime_1_dist}
    \end{subfigure}
    \caption{Results for the pattern $\mathit{Turn\cdot GapStart\cdot GapEnd\cdot Turn}$ with $m=1$.}\label{fig:real_maritime_1_contour}
\end{figure}

Figure \ref{fig:real_maritime_1_contour} shows results for the pattern 
\begin{equation}
\label{eq:turngap}
\mathit{Turn} \cdot \mathit{GapStart} \cdot \mathit{GapEnd} \cdot \mathit{Turn}
\end{equation}
where $\mathit{Turn}$ is shorthand notation for
\begin{equation*}
(\mathit{TurnNorth} + \mathit{TurnEast} + \mathit{TurnSouth} + \mathit{TurnWest})
\end{equation*} 
with $+$ denoting the $\mathit{OR}$ operator. 
Similarly for $\mathit{GapStart}$ and $\mathit{GapEnd}$.
With this pattern, we would like to detect a sequence of movements in which a vessel first turns
(regardless of heading),
then turns off its AIS equipment and subsequently re-appears by turning again.
Communication gaps are important for maritime analysts because they often indicate an intention of hiding
(e.g., in cases of illegal fishing in a protected area).
The aggregate precision score (Figure \ref{fig:real_maritime_1_acc_all}) is very close to the baseline performance.
A look at the per-state plots reveals something interesting
(Figures \ref{fig:real_maritime_1_acc}, \ref{fig:real_maritime_1_spread},  \ref{fig:real_maritime_1_dist}).
Note that, in order to avoid cluttering, 
we have removed duplicate states from the per-state plots.
In addition,
the superscript of each state in the $x$ axis shows the last event seen when in that state.
For example, the superscript $te$ corresponds to $\mathit{TurnEast}$, $tw$ to $\mathit{TurnWest}$, $tn$ to $\mathit{TurnNorth}$ and
$ts$ to $\mathit{TurnSouth}$ (states 3, 7, 9 and 11 respectively). Similarly for $\mathit{GapStart}$ for which superscripts start with $gs$ (states 13--16) and for $\mathit{GapEnd}$ ($ge$ and states 17--20).
%However, 
These per-state plots show that there is a distinct ``cluster'' of states (13--17) which exhibit high precision scores 
for all values of \pfc\ (Figure \ref{fig:real_maritime_1_acc}) and small spread for most values of \pfc\ (Figure \ref{fig:real_maritime_1_spread}).
Therefore, these states constitute what might be called ``milestones''
and a PMC can help in uncovering them. 
By closer inspection,
it is revealed that states 13--16 are visited after the PMC has seen one of the $\mathit{GapStart}$ events 
(we remind that $\mathit{GapStart}$ is a disjunction of the four directional sub-cases).
Moreover, $\mathit{GapEnd}$ events are very likely to appear in the input stream right after a $\mathit{GapStart}$
event, 
as expected, 
since during a communication gap (delimited by a $\mathit{GapStart}$ and a $\mathit{GapEnd}$),
a vessel does not emit any messages.
State 17, which also has a similar behavior, is visited after a $\mathit{GapEndNorth}$ event. 
Its high precision scores are due to the fact that, after a $\mathit{GapEnd}$ event,
a $\mathit{Turn}$ event is very likely to appear.
It differs from states 13--16 in its distance,
as shown in Figure \ref{fig:real_maritime_1_dist},
which is 1, whereas, for states 13--16, the distance is 2.
On the other hand, states 18--20,
which correspond to the other 3 $\mathit{GapEnd}$ events,
fail to produce any forecasts.
The reason is that there are no such $\mathit{GapEnd}$ events in the stream,
i.e., whenever this vessel starts transmitting again after a $\mathit{Gap}$,
it is always headed towards the northern direction. 

\begin{figure}[t]
\begin{centering}
\includegraphics[width=0.64\textwidth]{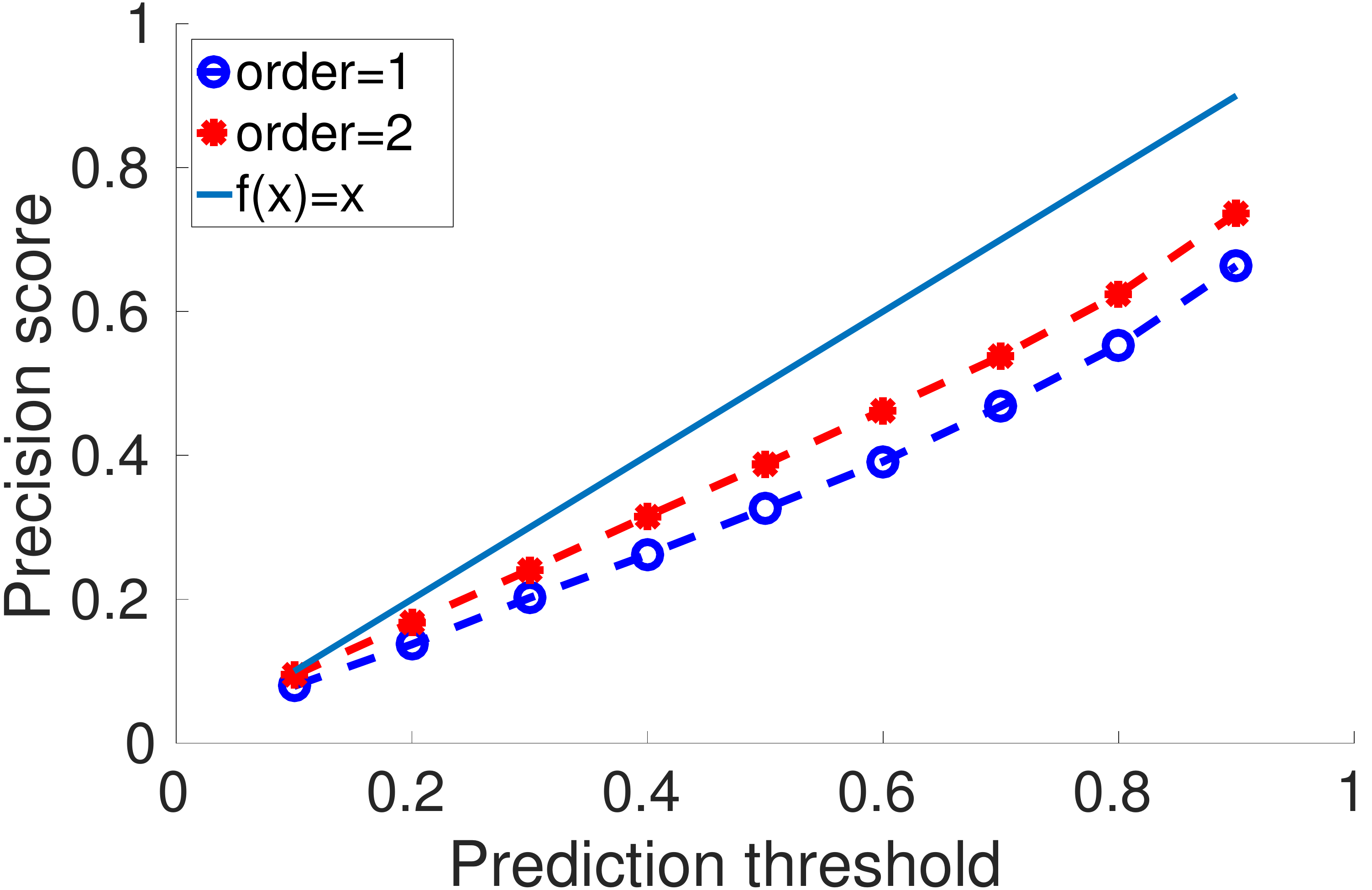}
\caption{Results for the pattern $\mathit{TurnNorth \cdot (TurnNorth + TurnEast)^{*} \cdot TurnSouth}$.}
\label{fig:real_maritime_northsouth_1_contour}
\end{centering}
\end{figure}

Figure \ref{fig:real_maritime_northsouth_1_contour} shows results for the pattern
\begin{equation*}
\mathit{TurnNorth} \cdot (\mathit{TurnNorth} + \mathit{TurnEast})^{*} \cdot \mathit{TurnSouth}
\end{equation*}
This pattern is more complex since it involves a \textit{star closure} operation on a nested \textit{union} operation.
It attempts to detect a rightward reverse of heading,
in which a vessel is initially heading towards the north and subsequently starts a right turn
until it ends up heading towards the south. 
Such patterns can be useful in detecting maneuvers of fishing vessels.

Figure \ref{fig:real_maritime_northsouth_1_contour} shows that a model with $m{=}1$ is unable to
approximate well-enough the correct waiting-time distribution.
Increasing the order to $m{=}2$ improves the precision score,
but it still remains under the baseline performance.
One could attempt to further increase the value of $m$,
but this would substantially increase the cost of building the PMC.
For $m=1$, the generated PMC has $\approx 30$ states.
For $m=2$, this number rises to $\approx 600$ and the cost of creating an unambiguous DFA and then its corresponding PMC rises exponentially.
When stationarity is assumed (as in our case) and the model does not need to be updated online,
an expensive model can be tolerated.

\subsection{Commentary on throughput}

So far, we have focused on precision and quality metrics.
For online event forecasting,
throughput (defined as $\frac{\#\ of\ events\ consumed}{total\ execution\ time}$, 
where \textit{execution time} refers to the \textbf{repeat} loop in Algorithm \ref{alg:wayeb}) is another important metric.
We omit presenting detailed results about throughput,
since Wayeb exhibits a steady behavior.
For the maritime use case and the more complex heading reversal pattern,
throughput is $\approx 1.2\times 10^{6}$ $\mathit{events/sec}$ and remains
steady for both $m{=}1$ and $m{=}2$,
whereas the event rate of the input stream is much lower.
The experiments for the maritime use case were run on a 64-bit Debian machine,
with Intel(R) Core(TM) i7-4770 CPU @ 3.40GHz processors, and 16GB of memory.
This high throughput number is due to the fact that the online operations of Wayeb consist mostly of memory operations (see Section \ref{sec:implementation}).
Even when the size of the PMC grows from $\approx 30$ to $\approx 600$,
there is minimal overhead in accessing and maintaining the larger look-up-tables of the latter PMC.
This independence from $m$ also holds when multiple runs are employed,
as in the credit card fraud use case 
(for the $\mathit{partition-contiguity}$ selection strategy).
Even in this case, only a single PMC is created (therefore, only one table for the DFA and one for the forecasts) and the different runs simply consult this PMC through a reference to it.
Throughput for the $\mathit{IncreasingAmounts}$ fraud pattern
is $\approx 1.2\times 10^{5}$ $\mathit{events/sec}$
(in total, 3 different patterns were tested),
whereas the event rate at peak times reaches up to $\approx 1000$ $\mathit{events/sec}$.
Due to privacy reasons, experiments on the fraud dataset were run in Feedzai's premises and thus on different hardware: a 64-bit Ubuntu machine, with Intel(R) Core(TM) i7-3770 CPU @ 3.40GHz processors and 32GB of memory.
The multiple runs that need to be created, accessed and maintained with this dataset (on the contrary, for the maritime use case, only a single run is created)
incur a significant increase in the execution time.

\section{Summary \& Future Work}
\label{sec:summary}

We presented Wayeb, a system that can produce online forecasts of event patterns.
This system is not restricted to sequential patterns,
but can handle patterns defined as regular expressions. 
It is also probabilistic and its forecasts can have guaranteed precision scores,
if the input stream is generated by a Markov process.
We have shown that it can provide useful forecasts even in real-world scenarios
in which we do not know beforehand the statistical properties of the input stream.
Moreover, the trade-off between precision score and the quality of the produced forecasts
has been explored.
Wayeb can also be used to uncover interesting probabilistic dependencies among the events involved in a pattern (pattern ``milestones''), 
which can be informative in themselves or could possibly be used for optimization purposes
in algorithms based on frequency statistics \cite{kolchinsky_lazy_2015}.

There are several 
%shortcomings which should be overcome 
%and that provide 
directions for future research.
One of them concerns relationality,
i.e., our system should be able to handle directly constraints between the arguments of different events within a pattern.
%One of them concerns the expressivity of the event patterns.
%In Complex Event Processing,
%patterns are usually defined as having constraints between the arguments 
%of the events involved. 
%The most usual is the $\mathit{within}$ constraint,
%requiring that all events should occur within a specified time window 
%(essentially this is a constraint involving the timestamps of the first and the last events in the sequence).
%Our system should be able to handle such constraints.
In this paper we focused on the \textit{non-overlap} counting policy and the \textit{contiguity} selection strategy. 
We have also implemented the \textit{overlap} policy, but did not discuss it due to space limitations.
With respect to the more flexible selection strategies
(like \textit{skip-till-any-match}),
the usual way to deal with them is to clone runs of the automaton online, when appropriate.
We could have followed a similar cloning approach as well and produce forecasts for each run.
However, it is doubtful whether individual forecasts made by a multitude (possibly hundreds)
of concurrently existing runs would be useful to a user. 
Some form of aggregate forecasting (e.g., number of full matches expected within the next $N$ events) could be more informative. 
We intend to pursue this line of research.
%As far as model learning is concerned,
%Wayeb can learn the transition matrix,
%but it cannot currently estimate the appropriate order of the PMC.
%This is a task that should also be automated.
Another useful functionality would be that of assessing whether the model should be updated online,
once we drop the stationarity assumption,
and how this could be done efficiently.

\bibliography{forecasting}

\end{document}